\documentclass[%
reprint,
superscriptaddress,
nofootinbib,
amsmath,amssymb,
aps,
prl,
]{revtex4-1}

\usepackage{graphicx} 
\usepackage{dcolumn}
\usepackage{bm}
\usepackage{braket}
\usepackage{dsfont}
\usepackage{color,colortbl}
\usepackage[table,xcdraw]{xcolor}
\usepackage{hyperref}
\usepackage{multirow}
\usepackage{subcaption}
\usepackage{mwe}
\usepackage{booktabs}
\usepackage{caption}
\usepackage{lipsum}
\usepackage{babel,blindtext}
\usepackage{amsmath}
\usepackage[toc,page]{appendix}
\usepackage[symbol*]{footmisc}
\usepackage{float}

\begin{document}
\title{Learning Local Equivariant Representations for Large-Scale Atomistic Dynamics}

\author{Albert Musaelian$^*$}
\affiliation{Harvard University}

\author{Simon Batzner$^{*, \dagger}$}
\affiliation{Harvard University}

\author{\\Anders Johansson}
\affiliation{Harvard University}

\author{Lixin Sun}
\affiliation{Harvard University}

\author{Cameron J. Owen}
\affiliation{Harvard University}

\author{Mordechai Kornbluth}
\affiliation{Robert Bosch LLC Research and Technology Center}

\author{Boris Kozinsky$^{\dagger}$}
\affiliation{Harvard University}
\affiliation{Robert Bosch LLC Research and Technology Center}

\def\thefootnote{*}\footnotetext{\textbf{Equal Contribution. Order is random.}\\}\def\thefootnote{\arabic{footnote}}

\def\thefootnote{$\dagger$}\footnotetext{Corresponding authors\\S.B., E-mail: \url{batzner@g.harvard.edu}\\B.K., E-mail: \url{bkoz@seas.harvard.edu}\\ }\def\thefootnote{\arabic{footnote}}

\newcommand\bvec{\mathbf}
\newcommand{\mathsc}[1]{{\normalfont\textsc{#1}}}

\begin{abstract}
A simultaneously accurate and computationally efficient parametrization of the energy and atomic forces of molecules and materials is a long-standing goal in the natural sciences. In pursuit of this goal, neural message passing has lead to a paradigm shift by describing many-body correlations of atoms through iteratively passing messages along an atomistic graph. This propagation of information, however, makes parallel computation difficult and limits the length scales that can be studied. Strictly local descriptor-based methods, on the other hand, can scale to large systems but do not currently match the high accuracy observed with message passing approaches. This work introduces Allegro, a strictly local equivariant deep learning interatomic potential that simultaneously exhibits excellent accuracy and scalability of parallel computation. Allegro learns many-body functions of atomic coordinates using a series of tensor products of learned equivariant representations, but without relying on message passing. Allegro obtains improvements over state-of-the-art methods on the QM9 and revised MD-17 data sets. A single tensor product layer is shown to outperform existing deep message passing neural networks and transformers on the QM9 benchmark. Furthermore, Allegro displays remarkable generalization to out-of-distribution data. Molecular dynamics simulations based on Allegro recover structural and kinetic properties of an amorphous phosphate electrolyte in excellent agreement with first principles calculations. Finally, we demonstrate the parallel scaling of Allegro with a dynamics simulation of 100 million atoms.
\end{abstract}

\maketitle

\section{Introduction}

Molecular dynamics (MD) and Monte-Carlo (MC) simulation methods for the study of properties of molecules and materials are a core pillar of computational chemistry, materials science, and biology. Common to a diverse set of applications ranging from energy materials \cite{richards2016design} to protein folding \cite{lindorff2011fast} is the requirement that predictions of the potential energy and the atomic forces must be both accurate and computationally efficient to faithfully describe the evolution of complex systems over long time scales. While first-principles methods such as density functional theory (DFT), which explicitly treat the electrons of the system, provide an accurate and transferable description of the system, they exhibit poor scaling with system size and thus limit practical applications to small systems and short simulation times. Classical force-fields based on simple functions of atomic coordinates are able to scale to large systems and long time scales but are inherently limited in their fidelity and can yield unfaithful dynamics. Descriptions of the potential energy surface (PES) using machine learning (ML) have emerged as a promising approach to move past this trade-off \cite{blank1995neural, handley2009optimal, behler2007generalized, gaporiginalpaper, thompson2015spectral, shapeev2016moment, schnet_jcp, sgdml, physnet_jctc, drautz2019atomic, christensen2020fchl, klicpera2020directional, nequip, dcf, gnnff, xie2021bayesian, xie2022uncertainty, deepmd, vandermause2020fly, vandermause2021active, anderson2019cormorant, kovacs2021linear}. Machine learning interatomic potentials (MLIPs) aim to approximate a set of high-fidelity energy and force labels at improved computational efficiency that scales linearly with the number of atoms. A variety of different approaches have been proposed, from shallow neural networks and kernel-based approaches \cite{blank1995neural, behler2007generalized, handley2009optimal, gaporiginalpaper} to more recent methods based on deep learning \cite{deepmd, schnet_neurips, klicpera2020directional, qiao2021unite, nequip}. In particular, a class of MLIPs based on message passing neural networks (MPNNs) has shown remarkable accuracy \cite{schnet_jcp, nequip, klicpera2020directional, schutt2021equivariant, qiao2021unite, physnet_jctc}. In interatomic potentials based on MPNNs, an atomistic graph is induced by connecting with edges each atom (node) to all neighboring atoms inside a finite cutoff sphere surrounding the central atom. Information is then iteratively propagated along this graph, allowing MPNNs to learn many-body correlations and access non-local information outside of the local cutoff. This iterated propagation, however, leads to large receptive fields with many effective neighbors for each atom, which slows down parallel computation and limits the length scales accessible to message passing MLIPs. MLIPs using \emph{strictly local} descriptors such as Behler-Parrinello neural networks \cite{behler2007generalized}, GAP \cite{gaporiginalpaper}, SNAP \cite{thompson2015spectral}, DeepMD \cite{deepmd}, Moment Tensor Potentials \cite{shapeev2016moment}, or ACE \cite{drautz2019atomic} do not suffer from this obstacle due to their strict locality. As a result they can easily be parallelized across devices and have successfully been scaled to extremely large system sizes \cite{jia2020pushing, lu202186, guo2022extending, nguyen2021billion}. Approaches based on local atom-density based descriptors, however, have so far fallen behind in accuracy compared to state-of-the-art equivariant message passing interatomic potentials \cite{nequip}.\\

In this work, we present Allegro, an equivariant deep learning approach that retains the high accuracy of the recently proposed class of equivariant MPNNs \cite{nequip, qiao2021unite, schutt2021equivariant, satorras2021n, haghighatlari2021newtonnet, brandstetter2021geometric} while combining it with strict locality and thus the ability to scale to large systems. We demonstrate that Allegro not only obtains state-of-the-art accuracy on a series of different benchmarks but can also be parallelized across devices to access hundreds of millions of atoms. We further find that Allegro displays a high level of transferability to out-of-distribution data, significantly outperforming other local MLIPs, in particular including body-ordered approaches. Finally, we show that Allegro can faithfully recover structural and kinetic properties from molecular dynamics simulations of Li\textsubscript{3}PO\textsubscript{4}, a complex phosphate electrolyte.

\subsection{Related Work}
\subsubsection{Message Passing Interatomic Potentials}
Message passing neural networks (MPNNs) based on learned atomistic representations have recently gained popularity in atomistic machine learning due to advantages in accuracy compared to hand-crafted descriptors. Message passing interatomic potentials operate on an atomistic graph constructed by representing atoms as nodes and defining edges between atoms within a fixed cutoff distance of one another. Each node is then represented by a hidden state $\bvec{h}_i^t \in \mathbb{R}^c$ representing the state of atom $i$ at layer $t$, and edges are represented by edge feature $\bvec{e}_{ij}$, for which the interatomic distance $d_{ij}$ is often used. The message passing formalism can then be concisely described as \cite{gilmer2017neural}: 
\begin{equation}
    \bvec{m}_i^{t+1} = \sum_{j \in \mathcal{N}(i)} M_t(\bvec{h}_i^t, \bvec{h}_j^t, \bvec{e}_{ij})
\end{equation}

\begin{equation}
   \bvec{h}_i^{t+1} = U_t(\bvec{h}_i^t, \bvec{m}_i^{t+1})
\end{equation}
where $M_t$ and $U_t$ are an arbitrary message function and node update function, respectively. From this propagation mechanism, it is immediately apparent that as messages are communicated over a sequence of $t$ steps, the local receptive field of an atom $i$, i.e. the effective set of neighbors that contribute to the final state of atom $i$, increases approximately cubically with the effective cutoff radius $r_{c, e}$. In particular, given a MPNN with $N_\text{layer}$ message passing steps and local cutoff radius of $r_{c, l}$, the resulting effective cutoff is $r_{c, e} = N_\text{layer} r_{c, l}$. Information from all atoms inside this receptive field ends up on a central atom's state $\bvec{h}_i$ at the final layer of the network. Due to the cubic growth of the number of atoms inside the receptive field cutoff $r_{c, e}$, parallel computation can quickly become unmanageable, especially for extended periodic systems. As an illustrative example, we may take a structure of 64 molecules of liquid water at pressure $P=1 \operatorname{bar}$ and temperature $T=300 \operatorname{K}$. For a typical setting of $N_t=6$ message passing layers with a local cutoff of $r_{c, l} = 6$\AA \ this would result in an effective cutoff cutoff of $r_{c, e}=36$\AA. While each atom only has approximately 96 atoms in its local 6\AA\ environment (including the central atom), it has 20,834 atoms inside the extended 36 \AA\ environment. Due to the message passing mechanism, information from each of these atoms flow  into the current central atom. In a parallel scheme, each worker must have access to the high-dimensional feature vectors $\bvec{h}_i$ of all 20,834 nodes, while the strictly local scheme only needs to have access to approximately $6^3 = 216$ times fewer atoms' states. From this simple example it becomes obvious that massive improvements in memory consumption and therefore scalability can be obtained from strict locality in machine learning interatomic potentials. It should be noted that conventional message passing allows for the possibility, in principle, to capture long-range interactions (up to $r_{c, e}$) and can induce many-body correlations. The relative importance of these effects in describing molecules and materials is an open question, and one of the aims of this work is to explore whether many-body interactions can be efficiently captured without increasing the effective cutoff.

\subsubsection{Equivariant Neural Networks}

The physics of atomic systems is unchanged under the action of a number of geometric symmetries---rotation, inversion, and translation---which together comprise the Euclidean group $E(3)$ (rotation alone is $SO(3)$, and rotation and inversion together comprise $O(3)$). Scalar quantities such as the potential energy are \emph{invariant} to these symmetry group operations, while vector quantities such as the atomic forces are \emph{equivariant} to them and transform correspondingly when the atomic geometry is transformed. More formally, a function between vector spaces $f: X \to Y$ is equivariant to a group $G$ if
\begin{equation}
    \label{eqn:equivariance}
    f(D_X[g]x) = D_Y[g]f(x) \quad \forall g \in G, \forall x \in X
\end{equation}
where $D_X[g] \in GL(X)$ is the representation of the group element $g$ in the vector space $X$. The function $f$ is invariant if $D_Y[g]$ is the identity operator on $Y$: in this case the output is unchanged by the action of symmetry operations on the input $x$.\\

Most existing MLIPs guarantee the invariance of their predicted energies by acting only on invariant inputs. In invariant message passing interatomic potentials in particular, each atom's hidden latent space is a feature vector consisting solely of invariant scalars \cite{schnet_neurips}.
More recently, however, a class of models known as equivariant neural networks \cite{thomas2018tensor,weiler20183d,kondor2018n,kondor2018clebsch} have been developed which can act directly on non-invariant geometric inputs, such as displacement vectors, in a symmetry-respecting way. This is achieved by using only $E(3)$-equivariant operations, yielding a model whose internal features are \emph{equivariant} with respect to the Euclidean group. Building on these concepts, equivariant architectures have been explored for developing interatomic potential models. Notably, the NequIP model \cite{nequip}, followed by several other equivariant implementations \cite{haghighatlari2021newtonnet, schutt2021equivariant, qiao2021unite, tholke2022torchmd, brandstetter2021geometric}, demonstrated unprecedentedly low error on a large range of molecular and materials systems, accurately describes structural and kinetic properties of complex materials, and exhibits remarkable sample efficiency. In both the present work and in NequIP, the representation $D_X[g]$ of an operation $g \in O(3)$ on an internal feature space $X$ takes the form of a direct sum of irreducible representations (commonly referred to as irreps) of $O(3)$. This means that the feature vectors themselves are naturally divided into sections based on which irrep they correspond to, or equivalently, how those features transform under symmetry operations. The irreps of $O(3)$, and thus the features, are indexed by a rotation order $\ell \geq 0$ and a parity $p \in (-1, 1)$. A tensor that transforms according to the irrep $\ell, p$ is said to ``inhabit'' that irrep.\\

A key operation in such equivariant networks is the tensor product of representations, an equivariant operation that combines two tensors $x$ and $y$ with irreps $\ell_1, p_1$ and $\ell_2, p_2$ to give an output inhabiting an irrep $\ell_\text{out}, p_\text{out}$ satisfying $| \ell_1 - \ell_2 | \leq \ell_\text{out} \leq | \ell_1 + \ell_2 |$ and $p_\text{out} = p_1 p_2$:
\begin{equation}
    \label{eqn:tp}
    (\mathbf{x} \otimes \mathbf{y})_{\ell_\text{out}, m_\text{out}} = \sum_{m_1, m_2}{
        \begin{pmatrix}
            \ell_1 & \ell_2 & \ell_\text{out} \\
            m_1 & m_2 & m_\text{out} 
        \end{pmatrix}
        \mathbf{x}_{\ell_1, m_1} \mathbf{y}_{\ell_2, m_2}
    }
\end{equation}
where $\begin{pmatrix}
            \ell_1 & \ell_2 & \ell_\text{out} \\
            m_1 & m_2 & m_\text{out} 
        \end{pmatrix}$ is the Wigner $3j$ symbol.
Two key properties of the tensor product are that it is bilinear (linear in both $\mathbf{x}$ and $\mathbf{y}$) and that it combines tensors inhabiting different irreps in a symmetrically valid way. Many simple operations are encompassed by the tensor product, such as for example:
\begin{itemize}
    \item scalar-scalar multiplication: $(\ell_1=0, p_1=1), (\ell_2=0, p_2=1) \to (\ell_\text{out}=0, p_\text{out}=1)$
    \item vector dot product: $(\ell_1=1, p_1=-1), (\ell_2=1, p_2=-1) \to (\ell_\text{out}=0, p_\text{out}=1)$
    \item vector cross product, resulting in a pseudovector: $(\ell_1=1, p_1=-1), (\ell_2=1, p_2=-1) \to (\ell_\text{out}=1, p_\text{out}=1)$.
\end{itemize}

The message function $M_t(\bvec{h}_i^t, \bvec{h}_j^t, \bvec{e}_{ij})$ of the NequIP model, for example, uses this tensor product to define a message from atom $j$ to $i$ as a tensor product between equivariant features of the edge $ij$ and the equivariant features of the neighboring node $j$.

\subsubsection{Atom-density Representations}
In parallel to message passing interatomic potentials, the Atomic Cluster Expansion (ACE) has been developed as a unifying framework for various descriptor-based MLIPs \cite{drautz2019atomic}. ACE is a systematic scheme for representing local atomic environments in a body-ordered expansion. The coefficients of the expansion of a particular atomic environment serve as an invariant description of that environment. To expand a local atomic environment, the local atomic density is first projected onto a combination of radial basis functions $R$ and spherical harmonic angular basis functions $\vec{Y}$:
\begin{equation}
    A_{z n \ell} = \sum_{j \in \mathcal{N}(i) \text{ s.t. } z_j = z} R_{n  \ell}(r_{ij})\vec{Y}_\ell^m(\hat{r}_{ij})
\end{equation}
where $z$ runs over all atom species in the system, $z_j$ is the species of atom $j$, $\mathcal{N}(i)$ is the set of all atoms within the cutoff distance of atom $i$, also known as its ``neighborhood'', and the $n$ index runs over the radial basis functions. The $m$ index on $A$ is implicit.
The basis projection of body order $\nu + 1$ is then defined as:
\begin{align}
    \label{eqn:ace}
    B_{\substack{z_1, n_1}}^{(\nu=1)} &= A_{z_1 n_1 \ell_1=0} \\
    \label{eqn:ace-b2}
    B^{(\nu=2)}_{\substack{z_1, z_2, n_1, n_2 \\ \ell_1, \ell_2}} &=  A_{z_1 n_1 \ell_1} \otimes  A_{z_2 n_2 \ell_2} \\
    &... \\
    \label{eqn:acearbrder}
    B^{(\nu)}_{\substack{z_1...z_\nu, n_1...n_\nu \\ \ell_1...\ell_\nu }} &= \bigotimes_{\alpha=1,...,\nu}{ A_{z_\alpha n_\alpha \ell_\alpha} }.
\end{align}
Only tensor products outputting scalars---which are invariant, like the final target total energy---are retained here. For example, in equation \ref{eqn:ace-b2}, only tensor products combining basis functions inhabiting the same rotation order $\ell_1 = \ell_2$ can produce scalar outputs. The final energy is then fit as a linear model over all the scalars $B$ up to some chosen maximum body-order $\nu + 1$.\\

It is apparent from equation \ref{eqn:acearbrder} that a core bottleneck in the Atomic Cluster Expansion is the polynomial scaling of the computational cost of evaluating the $B$ terms with respect to the total number of two-body radial-chemical basis functions $N_\text{full-basis}$ as the body order $\nu+1$ increases: $\mathcal{O}(N_\text{full-basis}^{\nu})$. In the basic ACE descriptor given above, $N_\text{full-basis} = N_\text{basis} \times S$ is the number of radial basis functions times the number of species. Species embeddings have been proposed for ACE to remove the direct dependence on $S$ \cite{darby2021compressing}. It retains, however, the $\mathcal{O}(N_\text{full-basis}^{\nu})$ scaling in the dimension of the embedded basis $N_\text{full-basis}$. NequIP and some other existing equivariant neural networks avert this unfavorable scaling by only computing tensor products of a more limited set of combinations of input tensors.

\section{Results}
In the following, we describe the proposed method for learning high-dimensional potential energy surfaces using strictly local many-body equivariant representations. 

\subsection{Energy decomposition}

We start by decomposing the potential energy of a system into per-atom energies $E_i$, in line with previous approaches \cite{behler2007generalized, gaporiginalpaper, schnet_neurips}:
\begin{equation}
    \label{eqn:toteng}
    E_{\text{system}} = \sum_{i}^{N}{\sigma_{Z_i} E_i + \mu_{Z_i}}
\end{equation}
where $\sigma_{Z_i}$ and $\mu_{Z_i}$ are per-species scale and shift parameters, which may be trainable. Unlike standard practice in existing MLIPs, we further decompose the per-atom energy into a sum over \emph{pairwise} energies, indexed by the central atom and one of its neighbors
\begin{equation}
    \label{eqn:edgedecomp}
    E_{i} = \sum_{j \in \mathcal{N}(i)}{\sigma_{Z_i, Z_j} E_{ij}}    
\end{equation}
where $j$ ranges over the neighbors of atom $i$, and again one may optionally apply a per-species-pair scaling factor $\sigma_{Z_i, Z_j}$. It is important to note that while these pairwise energies are \emph{indexed} by the atom $i$ and its neighbor $j$, they may depend on all neighbors $k \in \mathcal{N}(i)$. Because $E_{ij}$ and $E_{ji}$ contribute to different site energies $E_i$ and $E_j$, respectively, we choose that they depend only on the environments of the corresponding central atoms. As a result and by design, $E_{ij} \neq E_{ji}$. Finally, the force acting on atom $i$, $\vec{F}_i$, is computed using autodifferentiation according to its definition as the negative gradient of the total energy with respect to the position of atom $i$:
\[
\vec{F}_i = -\nabla_i E_{\text{system}}
\]
which gives an energy-conserving force field.

\subsection{The Allegro model}
\begin{figure*}
    \centering
    \includegraphics[width=0.9\textwidth]{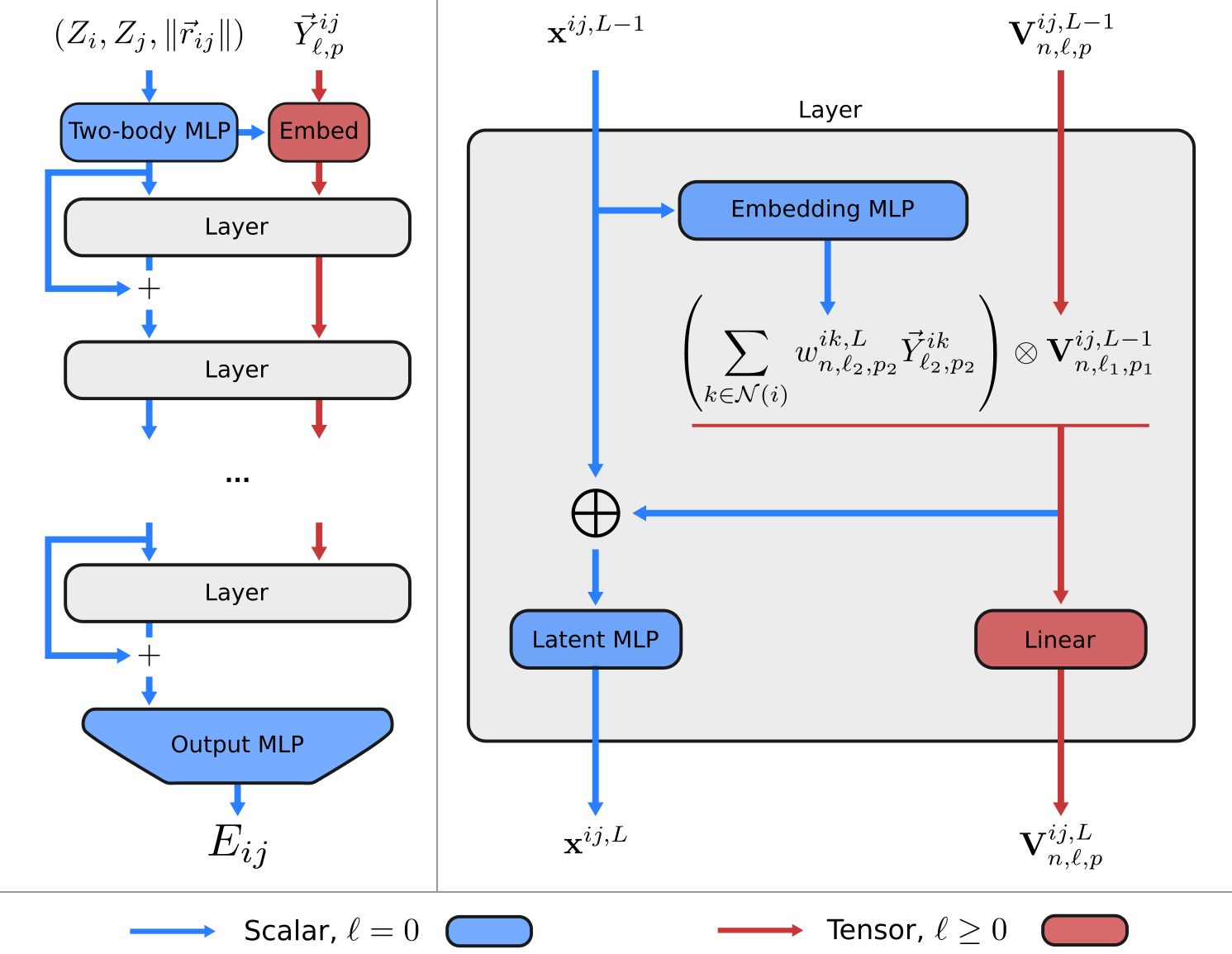}
    \caption{The Allegro network. Left: the Allegro model architecture. Right: a tensor product layer. Blue and red arrows represent scalar and tensor information, respectively, $\otimes$ denotes the tensor product, and $\oplus$ is concatenation.}
    \label{fig:Allegro}
\end{figure*}

The Allegro architecture, shown in figure \ref{fig:Allegro}, is an arbitrarily deep equivariant neural network with $N_{\text{layer}} \geq 1$ layers. The architecture learns representations associated with pairs of neighboring atoms using two latent spaces: an invariant latent space, which consists of scalar ($\ell=0$) features, and an equivariant latent space, which processes tensors of arbitrary rank $\ell \geq 0$. The two latent spaces interact with each other at every layer. The final pair energy $E_{ij}$ is then computed by a multi-layer perceptron (MLP) acting on the final layer's scalar features. \\

We use the following notations:
\begin{description}
    \item[$\vec{r}_i$] position of the $i$-th atom in the system
    \item[$\vec{r}_{ij}$]  relative displacement vector $\vec{r}_j - \vec{r}_i$ from $i$ to $j$
    \item[$r_{ij}$]  corresponding interatomic distance
    \item[$\hat{r}_{ij}$]  unit vector of $\vec{r}_{ij}$
    \item[$\vec{Y}^{ij}_{\ell,p}$] projection of $\hat{r}_{ij}$ onto the $\ell$-th real spherical harmonic which has parity $p = (-1)^\ell$. We omit the $m=-\ell,...,0,... \ell$ index within the representation from the notation for compactness
    \item[$Z_i$]  chemical species of atom $i$
    \item[$\mathrm{MLP}(...)$] multi-layer perceptron --- a fully-connected scalar neural network, possibly with nonlinearities
    \item[$\bvec{x}^{ij, L}$] invariant scalar latent features of the ordered pair of atoms $ij$ at layer $L$
    \item[$\bvec{V}^{ij, L}_{n,\ell,p}$] equivariant latent features of the ordered pair of atoms $ij$ at layer $L$. These transform according to a direct sum of irreps indexed by the rotation order $\ell \in 0, 1, ..., \ell_{\text{max}}$ and parity $p \in -1, 1$ and thus consist of both scalars ($\ell = 0$) and higher-order tensors ($\ell > 0$). The hyperparameter $\ell_{\text{max}}$ controls the maximum rotation order to which features in the network are truncated. In Allegro, $n$ denotes the channel index which runs over $0, ..., n_{\text{equivariant}} - 1$. We omit the $m$ index within each irreducible representation from the notation for compactness
\end{description}

\subsubsection{Two-body latent embedding}
\label{sec:initial-two-body}
Before the first tensor product layer, the scalar properties of the pair $ij$ are embedded through a nonlinear MLP to give the initial scalar latent features $\bvec{x}^{ij, L=0}$: 
\begin{equation}
\label{eqn:initial-scalars}
\bvec{x}^{ij, L=0} = \mathrm{MLP}_{\text{two-body}}\big( \mathsc{1Hot}(Z_i) \parallel \mathsc{1Hot}(Z_j) \parallel B(r_{ij}) \big)
\end{equation}
where $\parallel$ denotes concatenation, $\mathsc{1Hot}(\cdot)$ is a one-hot encoding of the center and neighbor atom species $Z_i$ and $Z_j$, and 
\begin{equation}
\label{eqn:radial-basis}
B(r_{ij}) = \big( B_1(r_{ij}) \parallel ... \parallel B_{N_\text{basis}}(r_{ij})) \big)
\end{equation}
is the projection of the interatomic distance $r_{ij}$ onto a radial basis. We use the Bessel basis function with a polynomial envelope function as proposed in \cite{klicpera2020directional}. The basis is normalized as described in \href{apx:norm-basis}{Appendix A} and plotted in Figure \ref{fig:bessel-basis}.\\

The initial \emph{equivariant} features $\bvec{V}^{ij, L=0}_{n,\ell,p}$ are computed as a linear embedding of the spherical harmonic projection of $\hat{r}_{ij}$:
\begin{equation}
\label{eqn:initial-equivariants}
\bvec{V}^{ij, L=0}_{n,\ell,p} = w^{ij,L=0}_{n,\ell,p} \vec{Y}^{ij}_{\ell, p}
\end{equation}
where the channel index $n = 0, ..., n_{\text{equivariant}} - 1$ and the scalar weights $w^{ij,L=0}_{n,\ell,p}$ for each center-neighbor pair $ij$ are computed from the initial two-body scalar latent features:
\begin{equation}
w^{ij,L=0}_{n,\ell,p} = \mathrm{MLP}_{\text{embed}}^{L=0}(\bvec{x}^{ij, L=0})_{n, \ell, p}.
\end{equation}
\\

\subsubsection{Layer architecture}

Each Allegro tensor product layer consists of four components:

\begin{enumerate}
    \item a MLP that generates weights to embed the central atom's environment
    \item an equivariant tensor product using those weights 
    \item a MLP to update the scalar latent space with scalar information resulting from the tensor product
    \item an equivariant linear layer that mixes channels in the equivariant latent space. \\
\end{enumerate} 

\paragraph{Tensor product} Our goal is to incorporate interactions between the current equivariant state of the center-neighbor pair and other neighbors in the environment, and the most natural operation with which to interact equivariant features is the tensor product. We thus define the updated equivariant features on the pair $ij$ as a weighted sum of the tensor products of the current features with the geometry of the various other neighbor pairs $ik$ in the local environment of $i$:
\begin{widetext}
\begin{align}
    \label{eq:tp}
    \bvec{V}^{ij, L}_{n,(\ell_1,p_1, \ell_2,p_2) \to (\ell_\text{out}, p_\text{out})} &= \sum_{k \in \mathcal{N}(i)}{ w_{n, \ell_2, p_2}^{ik,L}  \left( \bvec{V}^{ij, L-1}_{n,\ell_1,p_1} \otimes \vec{Y}^{ik}_{\ell_2,p_2} \right)
    } \\&= 
    \sum_{k \in \mathcal{N}(i)}{  \bvec{V}^{ij, L-1}_{n,\ell_1,p_1} \otimes \left( w_{n, \ell_2, p_2}^{ik,L}  \vec{Y}^{ik}_{\ell_2,p_2} \right)
    }
    \\&=
    \bvec{V}^{ij, L-1}_{n,\ell_1,p_1} \otimes \left( \sum_{k \in \mathcal{N}(i)}{w_{n, \ell_2, p_2}^{ik,L}  \vec{Y}^{ik}_{\ell_2,p_2}} \right)
\end{align}
\end{widetext}
where $k \in \mathcal{N}(i)$ ranges over the neighborhood $\mathcal{N}(i)$ of the central atom $i$. In the second and third lines we exploit the bilinearity of the tensor product in order to express the update in terms of \emph{one} tensor product, rather than one for each neighbor $k$, which saves significant computational effort. This is commonly referred to as the ``density trick'' \cite{gaporiginalpaper, bartok2013representing}.\\

Note that it is allowed for $(\ell_1, p_1) \neq (\ell_2, p_2) \neq (\ell_\text{out}, p_\text{out})$: valid tensor product paths are all those satisfying $| \ell_1 - \ell_2 | \leq  \ell_\text{out} \leq | \ell_1 + \ell_2|$ and $p_\text{out} = p_1 p_2$. We additionally enforce $\ell_\text{out} \leq \ell_\text{max}$. Which tensor product paths to include is a hyperparameter choice. In this work we include all allowable paths but other choices, such as restricting $(\ell_\text{out}, p_\text{out})$ to be among the values of $(\ell_1, p_1)$, are possible.\\

\paragraph{Environment embedding} The second argument to the tensor product, $\sum_{k \in \mathcal{N}(i)}{w_{n, \ell_2, p_2}^{ik,L} \vec{Y}^{ik}_{\ell_2, p_2}}$, is a weighted sum of the spherical harmonic projections of the various neighbor atoms in the local environment. This can be viewed as a weighted spherical harmonic basis projection of the atomic density, similar to the projection onto a spherical-radial basis used in ACE \cite{drautz2019atomic} and SOAP \cite{bartok2013representing}. For this reason, we refer to $\sum_{k \in \mathcal{N}(i)}{w_{n, \ell_2, p_2}^{ik,L} \vec{Y}^{ik}_{\ell_2, p_2}}$ as the ``embedded environment'' of atom $i$.\\

A central difference from the atomic density projections used in descriptor methods, however, is that the weights of the sum are learned. In descriptor approaches such as ACE, the $n$ index runs over a pre-determined set of radial-chemical basis functions, which means that the size of the basis must increase with both the number of species and the desired radial resolution. In Allegro, we instead leverage the previously learned scalar featurization of each center-neighbor pair to further learn
\begin{equation}
\label{eqn:env-embed-weights}
w_{n, \ell_2, p_2}^{ik,L} = \mathrm{MLP}_{\text{embed}}^{L}(\bvec{x}^{ik, L-1})_{n, \ell_2, p_2}    
\end{equation}
which yields an embedded environment with a fixed, chosen number of channels $n_\text{equivariant}$. It is important to note that $w_{n, \ell_2, p_2}^{ik,L}$ is learned as a function of the existing scalar latent representation of the center-neighbor pair $ik$ from previous layers. At later layers this can contain significantly more information about the environment of $i$ than a two-body radial basis. We typically choose $\mathrm{MLP}_{\text{embed}}$ to be a simple one-layer linear projection of the scalar latent space.\\

\paragraph{Latent MLP} Following the tensor product defined in \autoref{eq:tp}, the scalar outputs of the tensor product are reintroducted into the scalar latent space as follows: 
\begin{widetext}
\begin{equation}
\label{eqn:latent-mlp}
\bvec{x}^{ij, L} = \mathrm{MLP}_ {\text{latent}}^{L}\Big(
    \bvec{x}^{ij, L-1} \parallel
    \bigoplus_{(\ell_1,p_1, \ell_2,p_2)}{
        \bvec{V}^{ij, L}_{n,(\ell_1,p_1, \ell_2,p_2) \to (\ell_\text{out}=0, p_\text{out}=1)}
    }
\Big) \cdot u(r_{ij})
\end{equation}
\end{widetext}
where $\parallel$ denotes concatenation and $\oplus$ denotes concatenation over all tensor product paths whose outputs are scalars ($\ell_\text{out}=0, p_\text{out}=1$),  each of which contributes $n_{\text{equivariant}}$ scalars. The function $u(r_{ij}): \mathbb{R} \to \mathbb{R}$ by which the output is multiplied is the same smooth cutoff envelope function as used in the initial radial basis function of equation \ref{eqn:radial-basis}. The purpose of the latent MLP is to compress and integrate information from the tensor product, whatever its dimension, into the fixed dimension invariant latent space. This operation completes the coupling of the scalar and equivariant latent spaces since the scalars taken from the tensor product contain information about non-scalars previously only available to the equivariant latent space.\\

\paragraph{Mixing equivariant features}
Finally, the outputs of various tensor product paths with the same irrep $(\ell_\text{out}, p_\text{out})$ are linearly mixed to generate output equivariant features $\bvec{V}^{ij, L}_{n,\ell,p}$ with the same number of channels indexed by $n$ as the input features had:
\begin{widetext}
\begin{equation}
\label{eqn:equivar-linear}
\bvec{V}^{ij, L}_{n,\ell,p} = \sum_{\substack{ n' \\ (\ell_1, p_1, \ell_2, p_2) } }{
    w^{L}_{n,n',(\ell_1,p_1, \ell_2,p_2) \to (\ell, p)}
    \bvec{V}^{ij, L}_{n',(\ell_1,p_1, \ell_2,p_2) \to (\ell, p)}
}.
\end{equation}
\end{widetext}

The weights $w^{L}_{n,n',(\ell_1,p_1, \ell_2,p_2) \to (\ell, p)}$ are learned. This operation compresses the equivariant information from various paths with the same output irrep $(\ell, p)$ into a single output space regardless of the number of paths.\\

We finally note that an $SE(3)$-equivariant version of Allegro, which is sometimes useful for computational efficiency, can be constructed identically to the $E(3)$-equivariant model described here by simply omitting all parity subscripts $p$.

\subsubsection{Residual update}
After each layer, Allegro uses a residual update \cite{he2016deep} in the scalar latent space that updates the previous scalar features from layer $L-1$ by adding the new features to them (see appendix B). The residual update allows the network to easily propagate scalar information from earlier layers forward.

\subsubsection{Output block}

Finally, to predict the pair energy $E_{ij}$, we apply a fully-connected neural network with output dimension $1$ to the latent features output by the final layer:
\begin{equation}
    \label{eq:output-block}
    E_{ij} = \mathrm{MLP}_{\text{output}}(\bvec{x}^{ij, L=N_{\text{layer}}})
\end{equation}

\subsection{Normalization}
\label{sec:normalization}

\subsubsection{Internal normalization}

The normalization of neural networks' internal features is known to be of great importance to training. In this work we follow the normalization scheme of the \texttt{e3nn} framework \cite{e3nn_docs}, in which the initial weight distributions and normalization constants are chosen so that all components of the network produce outputs that element-wise have approximately zero mean and unit variance. In particular, all sums over multiple features are normalized by dividing by the square root of the number of terms in the sum, which follows from the simplifying assumption that the terms are uncorrelated and thus that their variances add. Two consequences of this scheme that merit explicit mention are the normalization of the embedded environment and atomic energy. Both the embedded environment (equation \ref{eqn:tp}) and atomic energy (equation \ref{eqn:edgedecomp}) are sums over all neighbors of a central atom. Thus we divide both by $\sqrt{\langle | \mathcal{N}(i) | \rangle}$ where $\langle | \mathcal{N}(i) | \rangle$ is the average number of neighbors over all environments in the entire training data set.

\subsubsection{Normalization of targets}

We found the normalization of the targets, or equivalently the choice of final scale and shift parameters for the network's predictions (see equation \ref{eqn:toteng}), to be of high importance. For systems of fixed chemical composition, our default initialization is the following: $\mu_{Z}$ is set for all species $Z$ to the average per-atom potential energy over all training frames $\braket{\frac{E_\mathrm{config}}{N}}$; $\sigma_{Z}$ is set for all species $Z$ to the root-mean-square of the components of the forces on all atoms in the training data set. This scheme ensures size extensivity of the potential energy, which is required if one wants to evaluate the potential on systems of different size than what it was trained on. We note that the widely used normalization scheme of subtracting the mean total potential energy across the training set violates size extensivity.\\

For systems with \emph{varying} chemical composition, we found it helpful to normalize the targets using a linear pre-fitting scheme that explicitly takes into account the varying chemical compositions: $\mu_{Z}$ is computed by  $\left[ N_{\mathrm{config}, Z} \right ] ^{-1} \left[E_{\mathrm{config}}\right]$, where $\left[ N_{\mathrm{config}, Z} \right ]$ is a matrix containing the number of atoms of each species in the reference structures, and $\left[E_{\mathrm{config}}\right]$ is a vector of reference energies. Details of the normalization calculations and the comparison between different schemes can be found in \cite{lixin-init}.

\subsection{Small molecule dynamics}    

\begin{table*}[!htbp]
\centering
\resizebox{\textwidth}{!}{\begin{tabular}{llccccccccc}
\hline \hline
Molecule & & FCHL19 \cite{christensen2020fchl, christensen2020role}  & UNiTE \cite{qiao2021unite} & GAP \cite{gaporiginalpaper} & ANI-pretrained \cite{ani-1, ani-2} & ANI-random  \cite{ani-1, ani-2} & ACE \cite{drautz2019atomic} & GemNet-(T/Q) \cite{klicpera2021gemnet} & NequIP (l=3) \cite{nequip} & Allegro \\
\hline
\multirow{2}{*}{Aspirin}  & \textit{Energy} & 6.2 & 2.4 & 17.7 & 16.6 &   25.4   & 6.1 & -  &\textbf{ 2.3} & \textbf{2.3}\\ 
                          & \textit{Forces} & 20.9 & 7.6& 44.9 & 40.6 &   75.0   & 17.9 & 9.5  & 8.2 &\textbf{ 7.3} \\ \hline

\multirow{2}{*}{Azobenzene}  & \textit{Energy} & 2.8 & 1.1 & 8.5 & 15.9 & 19.0      & 3.6 & -  &\textbf{ 0.7} & 1.2 \\ 
                          & \textit{Forces} & 10.8 & 4.2 & 24.5 & 35.4 &  52.1    & 10.9 & -  &2.9  &\textbf{2.6 }\\  \hline
                          
\multirow{2}{*}{Benzene}  & \textit{Energy} & 0.3 & 0.07 & 0.75 & 3.3 &   3.4   & \textbf{ 0.04} & -  & \textbf{0.04}&   0.3\\ 
                          & \textit{Forces} & 2.6 & 0.73 & 6.0 & 10.0 & 17.5     & 0.5 & 0.5 &0.3  & \textbf{0.2}\\  \hline
                          
\multirow{2}{*}{Ethanol}  & \textit{Energy} & 0.9 & 0.62 & 3.5 & 2.5 &  7.7    & 1.2 & - &  \textbf{0.4} &\textbf{0.4} \\ 
                          & \textit{Forces} & 6.2 & 3.7 & 18.1 & 13.4 &   45.6   & 7.3 & 3.6 & 2.8  &\textbf{2.1} \\ \hline
                          
\multirow{2}{*}{Malonaldehyde}  & \textit{Energy} & 1.5 & 1.1 & 4.8 & 4.6 &   9.4   & 1.7 & - & 0.8  & \textbf{0.6}\\ 
                          & \textit{Forces} & 10.2 & 6.6 & 26.4 & 24.5 &   52.4    & 11.1 & 6.6 & 5.1 & \textbf{3.6} \\  \hline
                          
\multirow{2}{*}{Naphthalene}  & \textit{Energy} & 1.2 & 0.46 & 3.8 & 11.3 &    16.0   & 0.9 & - &\textbf{ 0.2} & 0.5\\ 
                          & \textit{Forces} & 6.5 & 2.6 & 16.5 & 29.2 &    52.2  & 5.1 & 1.9 & 1.3  & \textbf{0.9}\\ \hline

\multirow{2}{*}{Paracetamol}  & \textit{Energy} & 2.9 &  1.9 & 8.5 & 11.5 &   18.2   & 4.0  & - & \textbf{1.4 } &1.5 \\ 
                          & \textit{Forces} & 12.2 & 7.1  & 28.9 & 30.4 &    63.3   & 12.7 & - &5.9  & \textbf{4.9}\\  \hline

\multirow{2}{*}{Salicylic acid}  & \textit{Energy} & 1.8 & 0.73& 5.6 & 9.2 &   13.5     & 1.8 &  - & \textbf{0.7} &0.9 \\  
                          & \textit{Forces} & 9.5 & 3.8 & 24.7 & 29.7 &     52.0 & 9.3 & 5.3 & 4.0 &\textbf{2.9} \\  \hline
                          
\multirow{2}{*}{Toluene}  & \textit{Energy} & 1.6 & 0.45 & 4.0 & 7.7 &    12.6   & 1.1 & - & \textbf{0.3} & 0.4\\ 
                          & \textit{Forces} & 8.8 & 2.5 & 17.8 & 24.3 &  52.9     & 6.5 & 2.2 &\textbf{ 1.6 }& 1.8\\ \hline
                          
\multirow{2}{*}{Uracil}  & \textit{Energy} & 0.6 & 0.58 & 3.0 & 5.1 &  8.3    & 1.1 & - &  \textbf{0.4}& 0.6\\ 
                          & \textit{Forces} & 4.2 & 3.8 & 17.6 & 21.4&   44.1    & 6.6 & 3.8 & 3.1 & \textbf{1.8}\\ \hline
  \hline \hline
    \end{tabular}}
    \caption{MAE on the revised MD-17 data set for energies and force components, in units of [meV] and [meV/{\AA}], respectively. Results for GAP, ANI, and ACE as reported in \cite{kovacs2021linear}. Best results are marked in \textbf{bold}. ANI-pretrained refers to a version of ANI that was pre-trained on 8.9 million structures and fine-tuned on the revMD17 data set, ANI-random refers to a randomly initialized model trained from scratch.}
    \label{tab:revmd17}
\end{table*}

We benchmark Allegro's ability to accurately learn energies and forces of small molecules on the revised MD-17 data set \cite{christensen2020role}, a recomputed version of the original MD-17 data set \cite{chmiela2017machine, schutt2017quantum, sgdml} that contains 10 small, organic molecules at DFT accuracy. As shown in table \ref{tab:revmd17}, we find that Allegro obtains state-of-the-art accuracy in the mean absolute error (MAE) in force components, while on some molecules, NequIP performs better on the energies. The authors note that while an older version of the MD-17 data set which has widely been used to benchmark MLIPs exists \cite{chmiela2017machine, schutt2017quantum, sgdml}, it has been observed to contain noisy labels \cite{christensen2020role} and is thus only of limited use for comparing the accuracy of MLIPs. 

\subsection{Transferability to higher temperatures}
\label{sec:3bpa}

\begin{table*}[!htbp]
\centering
\resizebox{\textwidth}{!}{\begin{tabular}{lcccccc|cc}
\hline \hline
            & ACE  \cite{drautz2019atomic} & sGDML \cite{sgdml}    & GAP \cite{gaporiginalpaper} & FF \cite{gaff-1, gaff-2}        & ANI-pretrained \cite{ani-1, ani-2}   & ANI-2x \cite{ani-1, ani-2}    & NequIP \cite{nequip}            & Allegro \\
\hline 
\textbf{Fit to 300K } \\ \hline 
300 K, E    & 7.1  & 9.1        & 22.8  & 60.8      & 23.5      &  38.6     &  \textbf{3.28 (0.10)} & 3.84 (0.08)   \\  
300 K, F    & 27.1 & 46.2       & 87.3  & 302.8     & 42.8      & 84.4      &  \textbf{10.77 (0.19)} & 12.98 (0.17)   \\     \hline
600 K, E    & 24.0 & 484.8      & 61.4  & 136.8     & 37.8      &  54.5     & \textbf{ 11.16 (0.14)} & 12.07 (0.45)   \\       
600 K, F    & 64.3 & 439.2      & 151.9 & 407.9     & 71.7      & 102.8     &  \textbf{26.37 (0.09)} & 29.11 (0.22)   \\       \hline 
1200 K, E   & 85.3 & 774.5      & 166.8 & 325.5     & 76.8      & 88.8      &  \textbf{38.52 (1.63)} & 42.57 (1.46)  \\             
1200 K, F   & 187.0  & 711.1    & 305.5 & 670.9     & 129.6     & 139.6     &  \textbf{76.18 (1.11)} & 82.96 (1.77)  \\  \hline 
\hline \hline
    \end{tabular}}
    \caption{Energy and Force RMSE for the 3BPA temperature transferability data set, reported in units of [meV] and [meV/{\AA}]. All models were trained on T=300K. Results for all models except for NequIP and Allegro from \cite{kovacs2021linear}. Best results are marked in \textbf{bold}. Models on the right of the vertical line are equivariant neural networks. For NequIP and Allegro we report the mean over 3 different seeds as well as the population standard deviation in parentheses.}
    \label{tab:3bpa}
\end{table*}

\begin{figure}
    \centering
    \includegraphics[width=0.95\linewidth]{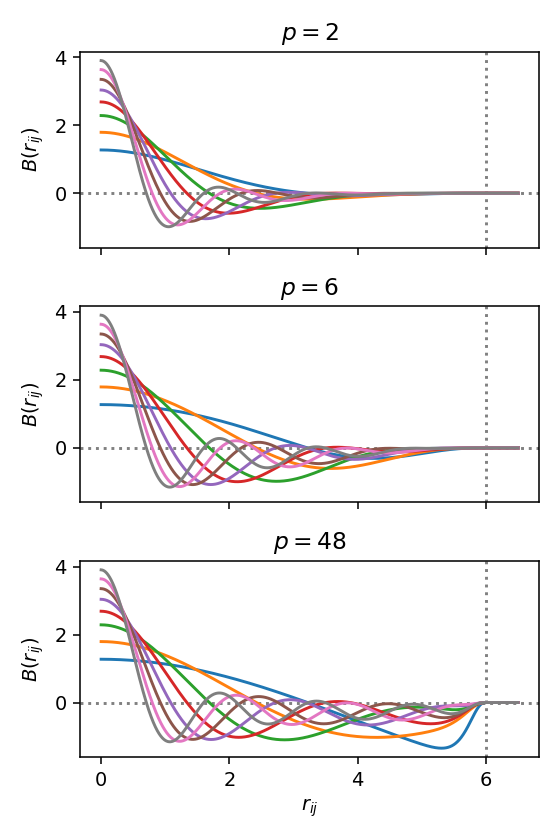}
    \caption{The normalized two-body radial basis functions $B_n(r_{ij})$ for $N_\text{basis} = 8$. The three panels show different values for the parameter $p$ of the smooth cutoff envelope. The cutoff $r_\text{max}$ is marked by the vertical line.}
    \label{fig:bessel-basis}
\end{figure}

For an interatomic potential to be useful in practice, it is crucial that it be transferable to new structures that might be visited over the course of a long molecular dynamics simulation. As a test of Allegro's generalization capabilities, we test the transferability to conformations sampled from higher-temperature MD simulations. We use the temperature transferability benchmark introduced in \cite{kovacs2021linear}: here, a series of data were computed using DFT for the flexible drug-like molecule 3-(benzyloxy)pyridin-2-amine (3PBA) at temperatures of 300K, 600K, and 1200K. A series of state-of-the-art methods were then benchmarked by training on 500 structures from the T=300K data set and then evaluating the trained models on data sampled at all three temperatures. Table \ref{tab:3bpa} shows a comparison of Allegro against existing approaches as they were reported in \cite{kovacs2021linear}. In particular, we compare against linear ACE \cite{drautz2019atomic}, sGDML \cite{sgdml}, GAP \cite{gaporiginalpaper}, a classical force-field based on the GAFF functional form \cite{gaff-1, gaff-2} as well as two ANI parametrizations \cite{ani-1, ani-2}: ANI-pretrained refers to a version of ANI that was pre-trained on 8.9 million structures and fine-tuned on this data set, while ANI-2x refers to the original parametrization trained on 8.9 million structures, but not fine-tuned on the 3BPA data set.  The equivariant neural networks Allegro and NequIP are seen to generalize significantly better than all other approaches. We found it particularly important to use a low exponent $p$ in the polynomial envelope function. We hypothesize that this is due to the fact that a lower exponent provides a stronger decay with increasing interatomic distance (see Figure \ref{fig:bessel-basis}), thereby inducing a stronger inductive bias that atoms $j$ further away from a central atom $i$ should have smaller pair energies $E_{ij}$ and thus contribute less to atom $i$'s site energy $E_i$. 

\subsection{Quantum-chemical properties of small molecules}

\begin{table*}[!htbp]
\centering
{\begin{tabular}{lllll}
\hline \hline
Model                                       & $U_0$       & $U$         & $H$         & $G$   \\ \hline
Schnet \cite{schnet_neurips}                & 14        & 19        & 14        & 14 \\
DimeNet++ \cite{klicpera2020fast}           & 6.3       & 6.3       & 6.5       & 7.6 \\
Cormorant \cite{anderson2019cormorant}      & 22        & 21        & 21        & 20 \\
LieConv \cite{finzi2020generalizing}        & 19        & 19        & 24        & 22 \\ 
L1Net   \cite{miller2020relevance}          & 13.5      & 13.8      & 14.4      & 14.0 \\ 
SphereNet \cite{liu2021spherical}           & 6.3        & 7.3       & 6.4       & 8.0 \\
EGNN  \cite{satorras2021n}                  & 11        & 12        & 12        & 12 \\
ET  \cite{tholke2022torchmd}                & 6.2        & 6.3      & 6.5       & 7.6 \\ 
NoisyNodes  \cite{godwin2021simple}         & 7.3       & 7.6       & 7.4       & 8.3 \\
PaiNN \cite{schutt2021equivariant}          & 5.9       & 5.7       & 6.0       & 7.4 \\ \hline
Allegro, 1 layer                            &  \underline{5.7} (0.2) &\underline{5.3} & \underline{5.3} & \underline{6.6}  \\
Allegro, 3 layers                           &  \textbf{4.7} (0.2) & \textbf{4.4} & \textbf{4.4} & \textbf{5.7}   \\
\hline \hline
\end{tabular}}
\caption{Comparison of models on the QM9 data set, measured by the MAE in units of [meV]. Allegro outperforms all existing message passing and transformer-based models, in particular even with a single layer. Best methods in bold, second-best method underlined.}
\label{tab:qm9}
\end{table*}

Next, we assess Allegro's ability to accurately model properties of small molecules across chemical space on the widely used QM9 data set \cite{ramakrishnan2014quantum}. The QM9 data set contains approximately 134k minimum-energy structures with chemical elements (C, H, O, N F) containing up to 9 heavy atoms (C, O, N, F) together with a series of corresponding molecular properties computed with DFT. We benchmark Allegro on four energy-related targets, in particular: a) $U_0$, the internal energy of the system at $T=0\operatorname{K}$  b) $U$, the internal energy at $T=298.15\operatorname{K}$, c) $H$, the enthalpy at $T=298.15\operatorname{K}$, and d) $G$, the free energy at $T=298.15\operatorname{K}$. This test is in contrast with other experiments conducted in this work that probed \emph{conformational} degrees of freedom, while here, we assess the ability of Allegro to describe properties across \emph{compositional} degrees of freedom. Table \ref{tab:qm9} shows a comparison with a series of state-of-the-art methods that also learn the properties described above as a direct mapping from atomic coordinates and species. We used 110,000 molecular structures for training, 10,000 for validation, and evaluated the test error on all remaining structures, in line with previous approaches \cite{schnet_jcp, schutt2021equivariant}. We note that Cormorant and EGNN are trained on 100,000 structures, L1Net is trained on 109,000 structures while NoisyNodes is trained on 114,000 structures. To give an estimate of variability of training as a function of random seed, we report for the $U_0$ target the mean and standard deviation across three different random seeds, resulting in different samples of training set as well as different weight initialization. We find that Allegro outperforms all existing methods. Surprisingly, even an Allegro model with a single tensor product layer obtains higher accuracy than all existing methods based on message passing neural networks and transformers. We note that the 1-layer and 3-layer Allegro networks have 7,375,237 and 17,926,533 parameters, respectively.

\subsection{Li-ion dynamics in an Phosphate Electrolyte}

\begin{figure*}[!htb]
\minipage{0.5\textwidth}
  \includegraphics[width=\linewidth]{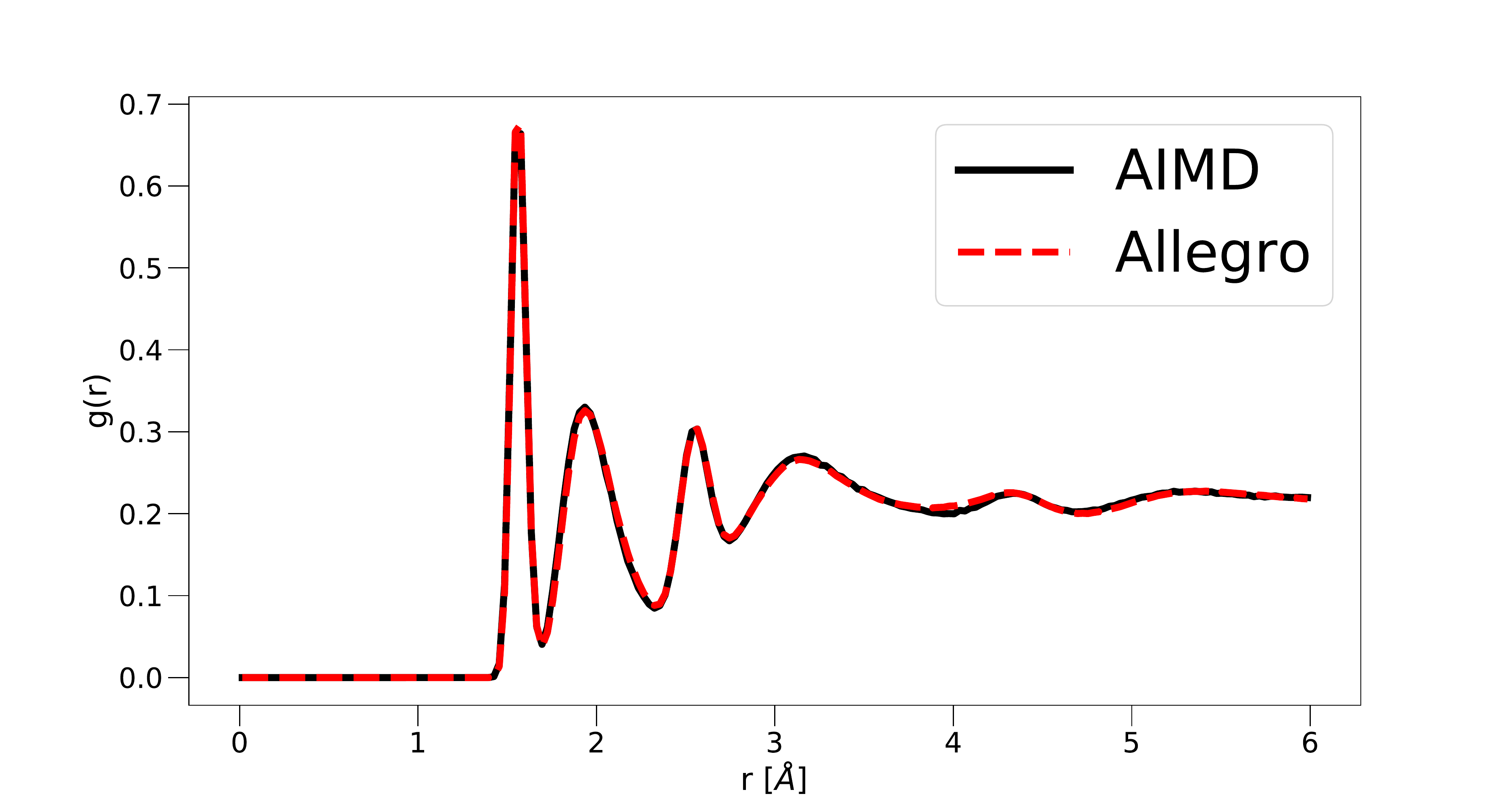}
\endminipage\hfill
\minipage{0.5\textwidth}
  \includegraphics[width=\linewidth]{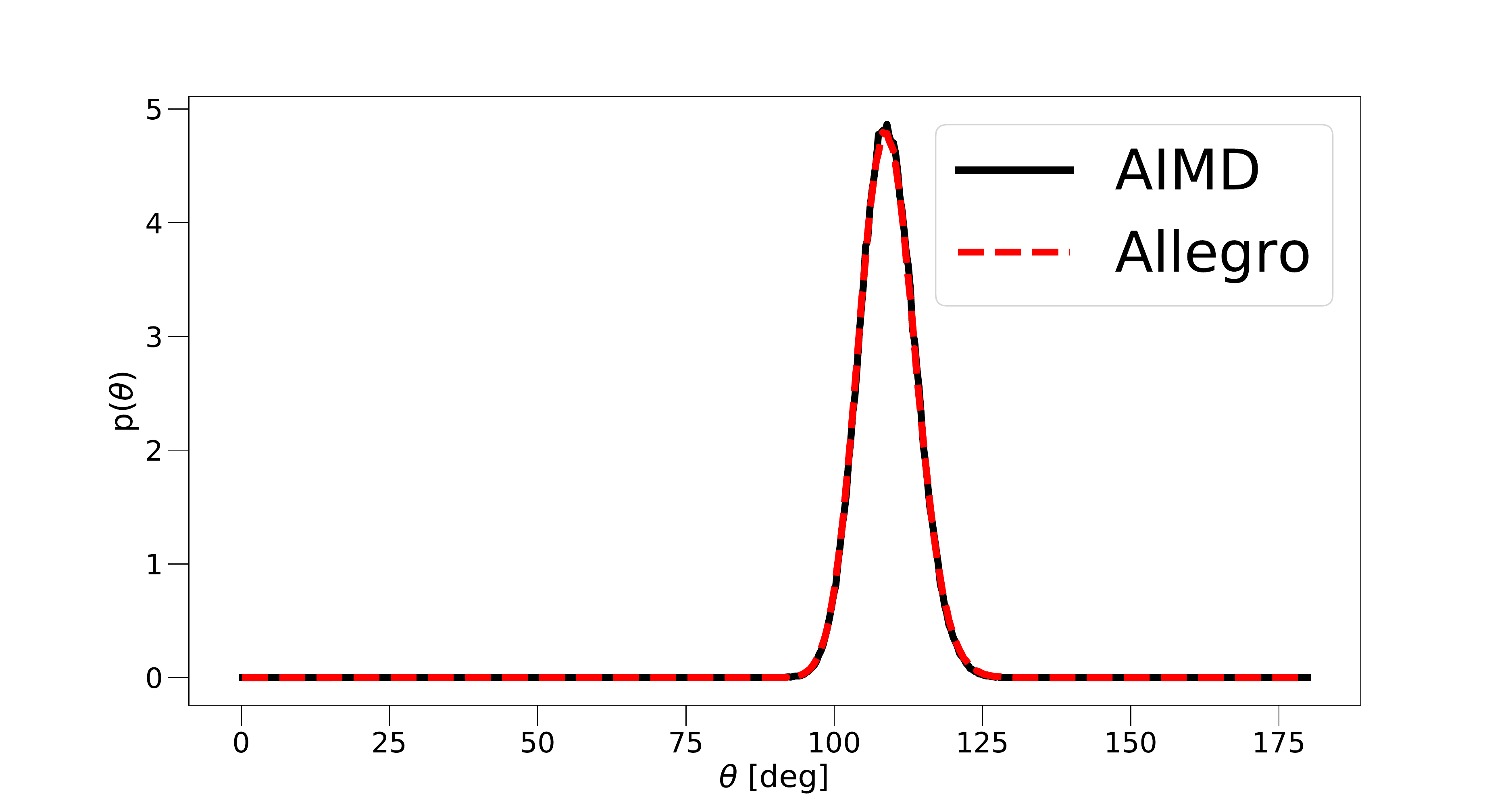}
\endminipage 
\caption{Structural Properties. Left: Radial Distribution Function, right:  Angular Distribution Function of tetrahedral bond angle. All defined as probability density functions.}
\label{fig:structure}
\end{figure*}

\begin{figure}[H]
    \centering
    \includegraphics[width=.75\columnwidth]{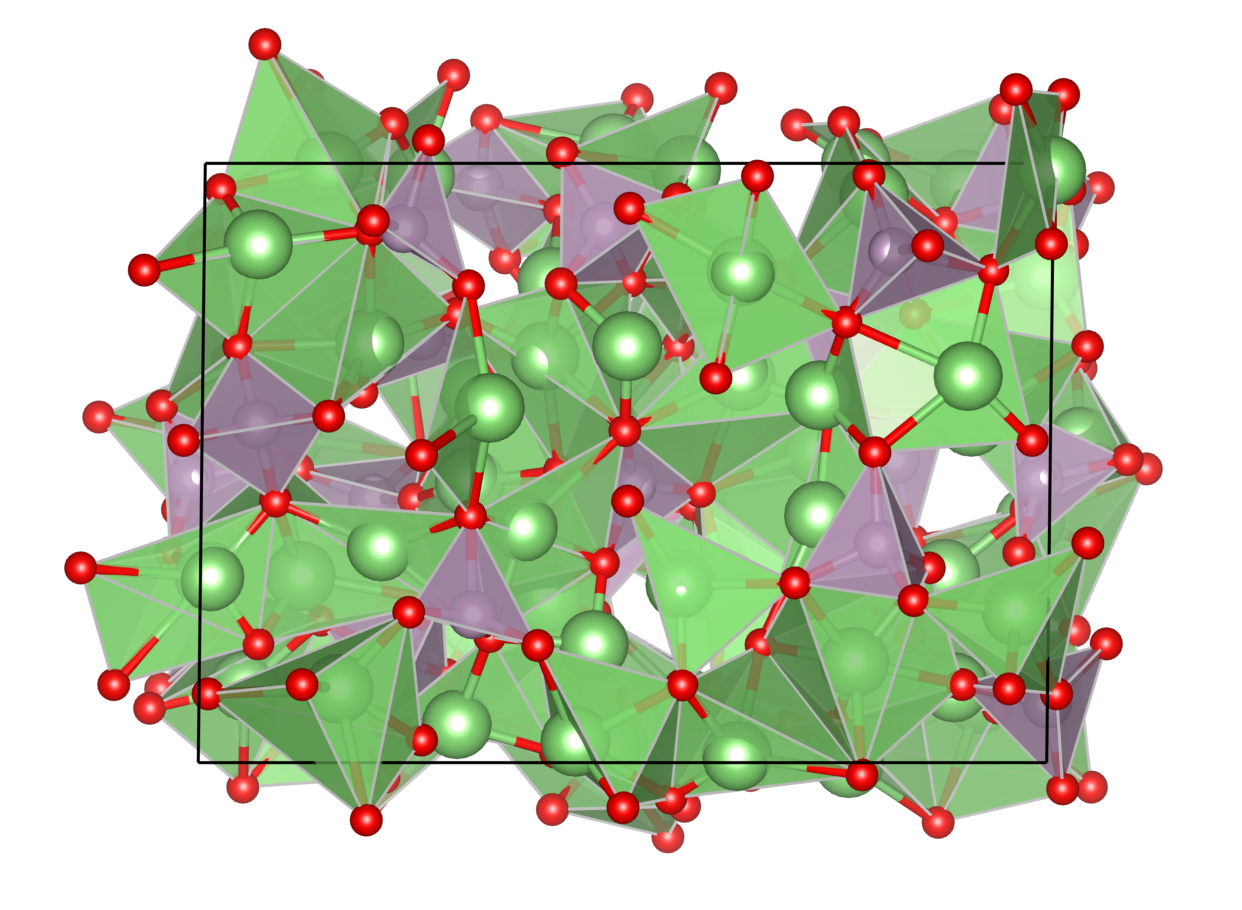}
    \caption{The quenched Li\textsubscript{3}PO\textsubscript{4} structure at T=600K}
    \label{fig:li3po4-quench}
\end{figure}

In order to examine Allegro's ability to describe kinetic properties from MD simulations, we use it to study amorphous structure formation and Li-ion migration in the Li\textsubscript{3}PO\textsubscript{4} solid electrolyte (see Figure \ref{fig:li3po4-quench}). This class of solid-state electrolytes is characterized by the intricate dependence of conductivity and mechanical properties on the degree of crystallinity \cite{LiPON,LiSiPON,Kalnaus2021,li2017study}. \\

\begin{figure}
    \centering
    \includegraphics[width=\columnwidth]{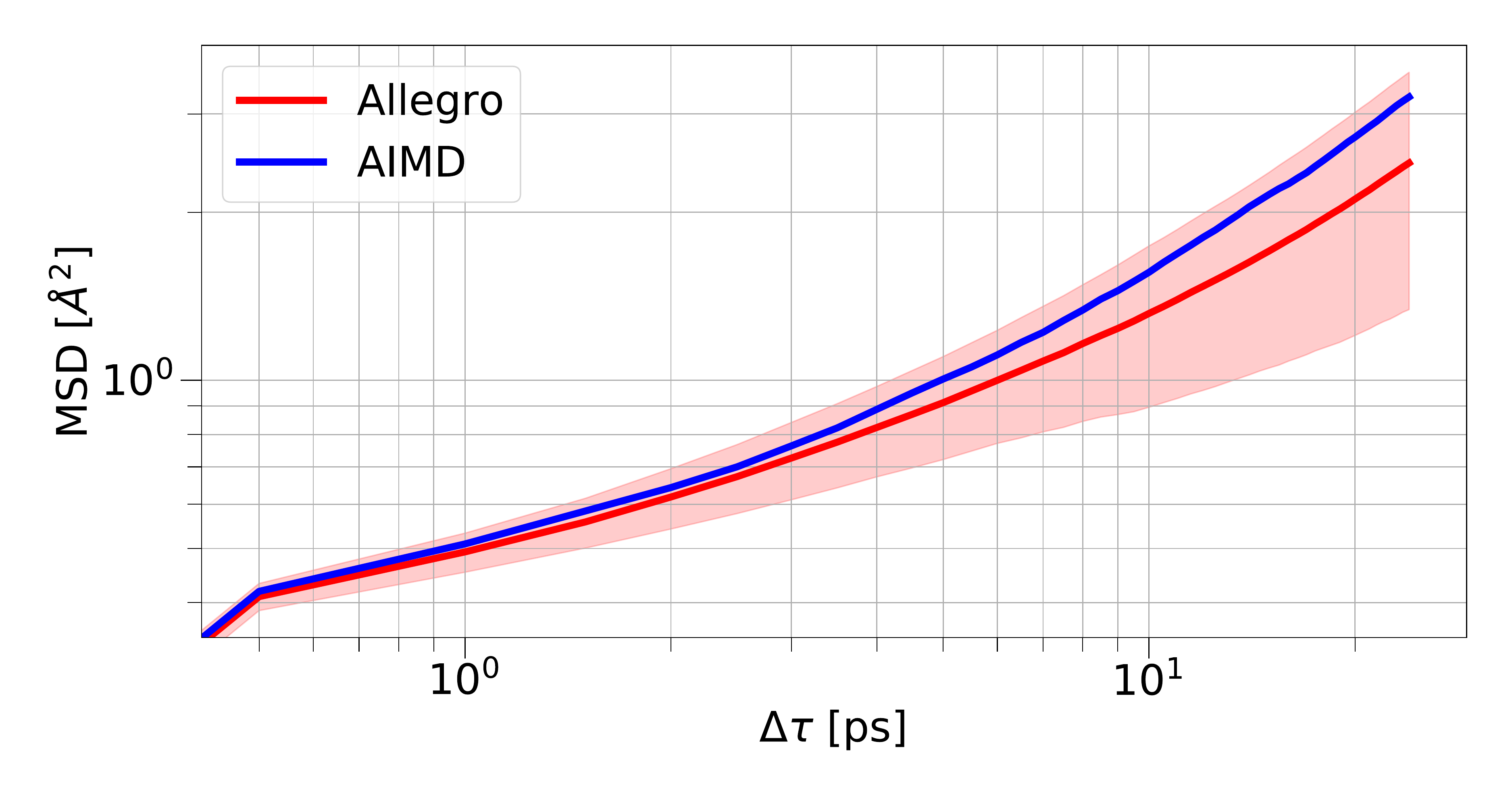}
    \caption{Li Dynamics. Comparison of the Li MSD of AIMD vs. Allegro. Results are averaged over 10 runs of Allegro, shading indicates +/- one standard deviation.}
    \label{fig:msd}
\end{figure}

In particular, the Li\textsubscript{3}PO\textsubscript{4} data set used in this work consists of two parts: a 50 ps \textit{ab-initio} molecular dynamics (AIMD) simulation in the melted state at T=3000K, followed by a 50 ps AIMD simulation in the quenched state at T=600K. We train a potential on structures from the melted and the quenched trajectories. The model used here is computationally efficient due to a relatively small number of parameters and tensor products and is identical to the one used in scaling experiments detailed below. When evaluated on the quenched amorphous state, which the simulation is performed on, a MAE in the energies of 1.7 meV/atom was obtained, as well as a MAE in the force components of 75.7 meV/\AA. We note that models of higher accuracy can be obtained by choosing larger networks, but we found the small, fast model to work sufficiently well to capture complex structural and kinetic properties under a phase change. We then run a series of ten MD simulations starting from the initial structure of the quenched AIMD simulation, all of length 50 ps at T=600K in the quenched state to compare how well Allegro recovers the structure and kinetics compared to AIMD. To assess the quality of the structure under the phase change, we compare the all-to-all radial distribution functions (RDF) and the angular distribution functions (ADF) of the tetrahedral angle P-O-O (P central atom). We show in Figure \ref{fig:structure} that Allegro can correctly recover both distribution functions. Next, we test how well Allegro can model the Li mean-square-displacement (MSD) in the quenched state. We again find excellent agreement with AIMD, as shown in Figure \ref{fig:msd}. 

\subsection{Scaling}

Many interesting phenomena in materials science, chemistry, and biology require large numbers of atoms, long time scales, a diversity of chemical elements, or often all three. Scaling to large numbers of atoms requires parallelization across multiple workers, which is difficult in MPNNs because the iterative propagation of atomic state information along the atomistic graph increases the size of the receptive field as a function of the number of layers. Allegro is designed to avoid this issue by strict locality and scales as
\begin{itemize}
    \item $\mathcal{O}(N)$ in the number of atoms in the system $N$, in contrast to the $\mathcal{O}(N^2)$ scaling of some global descriptor methods such as sGDML \cite{sgdml};
    \item $\mathcal{O}(M)$ in the number of neighbors per atom $M$, in contrast to the quadratic $\mathcal{O}(M^2)$ scaling of some deep learning approaches such as DimeNet \cite{klicpera2020directional} or Equivariant Transformers \cite{fuchs2020se, tholke2022torchmd};
    \item $\mathcal{O}(1)$ in the number of species $S$, unlike local descriptors such as SOAP ($\mathcal{O}(S^2)$) or ACE ($\mathcal{O}(S^{\text{body order} - 1})$) \cite{drautz2019atomic}.
\end{itemize}

In particular, in addition to scaling as $\mathcal{O}(N)$ in the number of atoms, Allegro is strictly local within the chosen cutoff and thus easy to parallelize. Recall that equations \ref{eqn:toteng} and \ref{eqn:edgedecomp} define the total energy of a system in Allegro as a sum over scaled pairwise energies $E_{ij}$. Thus by linearity, the force on atom $a$
\[
\vec{F}_a = -\nabla_a E_{\text{system}} = -\sum_{i,j}{\nabla_a E_{ij}}
\]
ignoring per-species scaling coefficients for clarity. Because each $E_{ij}$ depends only the atoms in the neighborhood of atom $i$, $-\nabla_a E_{ij} \neq 0$ only when $a$ is in the neighborhood of $i$. Further, for the same reason, pair energy terms $E_{ij}$ with different central atom indices $i$ are independent. As a result, these groups of terms may be computed independently for each central atom, which facilitates parallelization: the contributions to the force on atom $a$ due to the neighborhoods of various different atoms can be computed in parallel by whichever worker is currently assigned the relevant center's neighborhood. The final forces are then a simple sum reduction over force terms from various workers.\\

We first demonstrate the favorable scaling of Allegro in system size by parallelizing the method across GPUs on a single compute node as well as across multiple GPU nodes. The experiments were performed on NVIDIA DGX A100s on the Theta-GPU cluster at the Argonne Leadership Computing Facility, where each node contains 8 GPUs and a total of 320 GB of GPU memory. We choose two test systems for the scaling experiments: a) the quenched state structures of the multi-component electrolyte Li\textsubscript{3}PO\textsubscript{4} and b) the Ag bulk crystal with a vacancy, simulated at 90\% of the melting temperature. The Ag model used 1,000 structures for training and validation, resulting in energy MAE of 0.397 meV/atom and force MAE of 16.8 meV/\AA. Scaling numbers are dependent on a variety of hyperparameter choices, such as network size and radial cutoff, that control the trade-off between evaluation speed and accuracy. For Li\textsubscript{3}PO\textsubscript{4}, we explicitly choose these identically to those used in the previous set of experiments in order to demonstrate how well an Allegro potential scales that we demonstrated to give highly accurate prediction of structure and kinetics. We use the a time step of 2 fs, identical to the reference AIMD simulations, float32 precision, and a temperature of T=600K on the quenched structure, identical to the production simulations used above. For Ag, we use a time step of 5 fs, a temperature of T=300K and again float32 precision. Simulations were performed for 1,000 time steps after initial warm-up. Table \ref{tab:efficiency} show the computational efficiency on a series of system of varying size and computational resources. We are able to simulate the Ag system with over 100 million atoms on 16 GPU nodes.\\

\begin{table*}[!htbp]
\centering
{\begin{tabular}{lcccc}
\hline \hline
 Material & Number of atoms & Number of GPUs & Speed in ns/day & microseconds / (atom $\cdot$ step) \\ \hline \hline
 Li\textsubscript{3}PO\textsubscript{4} &  192        & 1         &      32.391        &   27.785  \\ \hline
 Li\textsubscript{3}PO\textsubscript{4} &  421,824     & 1         &     0.518        &   0.552    \\
 Li\textsubscript{3}PO\textsubscript{4} &  421,824   & 2         &       1.006       &     0.284   \\
 Li\textsubscript{3}PO\textsubscript{4} &  421,824    & 4         &     1.994        &    0.143    \\
 Li\textsubscript{3}PO\textsubscript{4} &  421,824     & 8         &    3.810         &     0.075    \\
 Li\textsubscript{3}PO\textsubscript{4} &  421,824    & 16        &     6.974         &      0.041  \\
 Li\textsubscript{3}PO\textsubscript{4} &  421,824    & 32        &      11.530        &    0.025       \\
 Li\textsubscript{3}PO\textsubscript{4} &  421,824    & 64        &      15.515      &     0.018    \\ \hline
 Li\textsubscript{3}PO\textsubscript{4} &  50,331,648        & 128       &    0.274         &     0.013      \\ \hline \hline
 Ag &  71              & 1     &     90.190        &     67.463     \\ \hline 
 Ag &  1,022,400       & 1     &       1.461      &      0.289    \\
 Ag &  1,022,400       & 2      &      2.648      &     0.160      \\
 Ag &  1,022,400       & 4      &      5.319     &          0.079  \\
 Ag &  1,022,400       & 8      &     10.180     &     0.042        \\
 Ag &  1,022,400       & 16      & 18.812     &          0.022     \\
 Ag &  1,022,400       & 32      &   28.156   &         0.015      \\
 Ag &  1,022,400       & 64      &   43.438   &        0.010       \\
 Ag &  1,022,400       & 128      &   49.395   &       0.009       \\ \hline
 Ag &  100,640,512      & 128      &    1.539  &           0.003   \\ \hline
\hline \hline
\end{tabular}}
\caption{Simulation times obtainable in [ns/day] and time required per atom per step in [microseconds] for varying number of atoms and computational resources.}
\label{tab:efficiency}
\end{table*}

\begin{figure}
    \centering
    \includegraphics[width=\columnwidth]{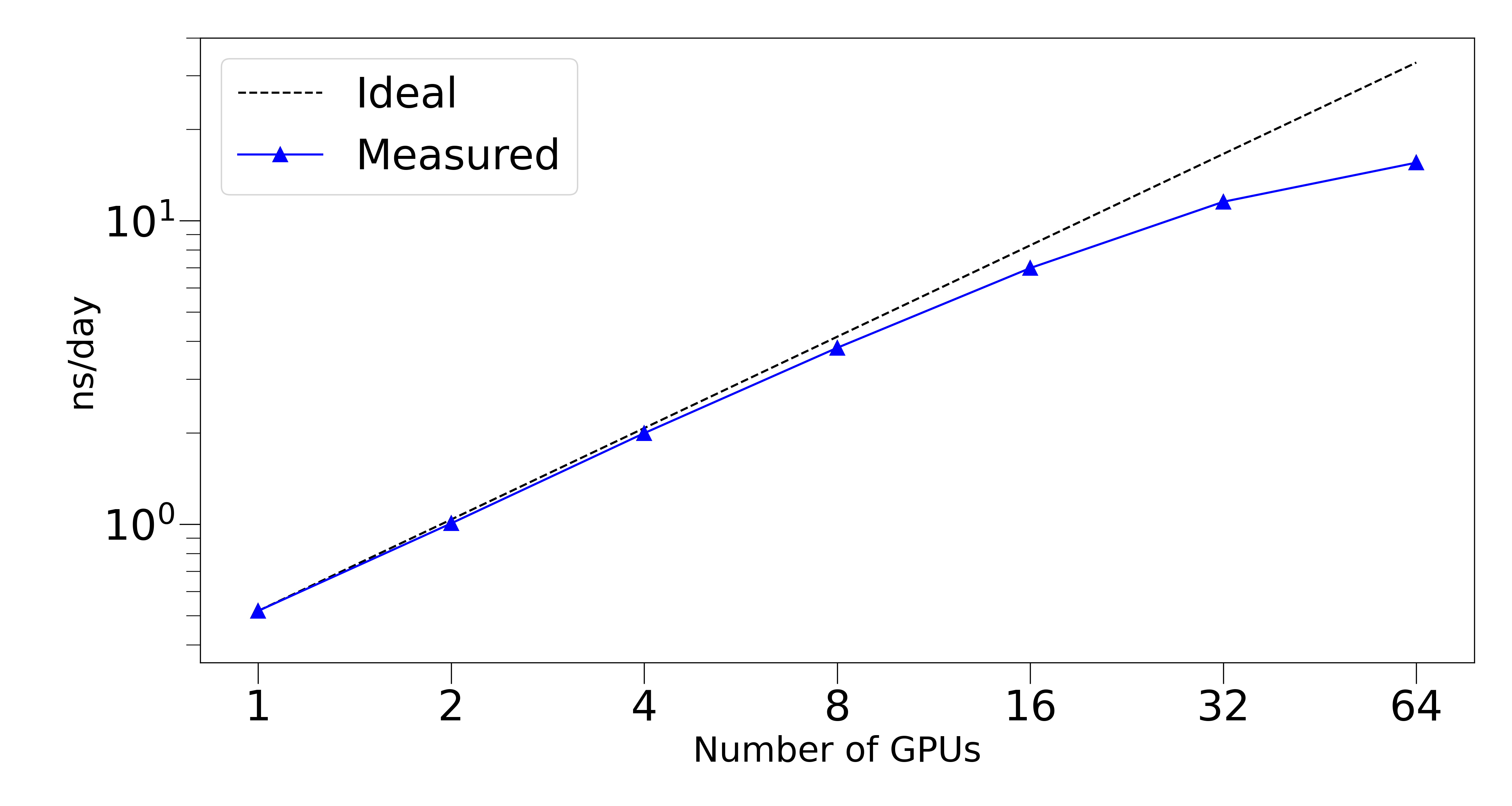}
    \caption{Strong Scaling results on a Li\textsubscript{3}PO\textsubscript{4} structure of 421,824 atoms, performed in LAMMPS.}
    \label{fig:strong-scaling}
\end{figure}

Scalability across devices is achieved by implementing an Allegro extension to the LAMMPS molecular dynamics code \cite{LAMMPS}. The local nature of the Allegro model is perfectly compatible with the spatial decomposition approach used in LAMMPS. All communication between MPI ranks is handled by existing LAMMPS functionality. The Allegro extension simply transforms the LAMMPS neighbor lists into the format required by the Allegro PyTorch model and stores the resulting forces and energies in the LAMMPS data structures. These operations are performed on the GPU by implementing an additional extension to LAMMPS's KOKKOS package, which uses the Kokkos performance portability library~\cite{kokkos} to entirely avoid expensive CPU work or CPU-GPU data transfer. This parallel architecture also allows multiple GPUs to be used to increase the speed of the potential calculation for a \emph{fixed} size system. Figure \ref{fig:strong-scaling} shows such strong scaling results on a $421,824$ atom Li\textsubscript{3}PO\textsubscript{4} structure. The system size was kept constant while the number of A100 GPUs used ranged over $\{1, 2, 4, 8, 16, 32, 64\}$.  

\subsection{Theoretical Analysis}

In this section, we provide a theoretical analysis of the suggested method. We do so by highlighting similarities and differences to the Atomic Cluster Expansion (ACE) framework \cite{drautz2019atomic}. Throughout this section we omit representation indices $\ell$ and $p$ from the notation for conciseness: every weight or feature that carries $\ell$ and $p$ indices previously implicitly carries them in this section. Starting from the initial equivariant features for the pair of atoms $ij$ at layer $L=0$
\begin{equation}
\bvec{V}^{ij, L=0}_{n_0} = w^{ij,L=0}_{n_0} \vec{Y}^{ij}
\end{equation}
the first Allegro layer computes a sum over tensor products between $\bvec{V}^{ij, L=0}_{n_0}$ and the spherical harmonics projection of all neighbors $k \in \mathcal{N}(i)$:
\begin{widetext}
\begin{align}
    \bvec{V}_{n_1}^{ij, L=1} &= \sum_{\substack{ n_1' \\ \text{paths} }}{
        w^{L=1}_{n_1,n_1',\text{path}}
        \sum_{k_1 \in \mathcal{N}(i)}{
            w_{n_1'}^{ik_1,L=1}
            \big( w_{n_1'}^{ij,L=0}\vec{Y}^{ij} \otimes \vec{Y}^{ik_1} \big)
        }
    }
    \\&=
    \label{eqn:one-layer-expanded}
    \sum_{\substack{ n_1' \\ \text{paths}}}{
        w^{L=1}_{n_1,n_1',\text{path}}
        \sum_{k_1 \in \mathcal{N}(i)}{
            w_{n_1'}^{ik_1,L=1}
            w_{n_1'}^{ij,L=0}
            \big( \vec{Y}^{ij} \otimes \vec{Y}^{ik_1} \big)
        }
    }
\end{align}
\end{widetext}
which follows from the bilinearity of the tensor product.
The sum over ``paths'' in this equation indicates the sum over all symmetrically valid combinations of implicit irrep indices on the various tensors present in the equation as written out explicitly in equation \ref{eqn:equivar-linear}.
Repeating this substitution, we can express the equivariant features at layer $L=2$ and reveal a general recursive relationship:
\begin{widetext}
\begin{align}
    \bvec{V}_{n_2,\ell_2,p_2}^{ij, L=2} &= \sum_{\substack{ n_2' \\ \text{paths}}}{
        w^{L=2}_{n_2,n_2',\text{path}}
        \sum_{k_2 \in \mathcal{N}(i)}{
            w_{n_2'}^{ik_2,L=2}
            \big( \bvec{V}_{n_2'}^{ij, L=1} \otimes \vec{Y}^{ik_2} \big)
        }
    }
    \\ &= 
    \label{eqn:unwrapped-two-layer}
    \sum_{\substack{ n_1', n_2' \\ \text{paths} }}{
        w^{L=2}_{n_2,n_2',\text{path}}
        w^{L=1}_{n_2',n_1',\text{path}}
        \left[
        \sum_{k_2 \in \mathcal{N}(i)}{ \sum_{k_1 \in \mathcal{N}(i)}{
            w_{n_2'}^{ik_2,L=2}
            w_{n_1'}^{ik_1,L=1}
            w_{n_1'}^{ij,L=0}
            \big( \vec{Y}^{ij} \otimes \vec{Y}^{ik_1} \otimes \vec{Y}^{ik_2} \big)
        }}
        \right]
    }
    \\
    \label{eqn:unwrapped-equiv-feat}
    \bvec{V}_{n_L, \ell_L, p_L}^{ij, L} &=
    \sum_{\substack{ k_1, ..., k_L \\ n_1', ..., n_L' \\ \text{paths} }}{
    \left[
        \left( \prod_{\alpha \in 1, ..., L}{ w^{L=\alpha}_{n_{\alpha + 1}',n_{\alpha}',\text{path}}} \right)
        \left( \prod_{\alpha \in 0, ..., L}{ w_{n_\alpha'}^{ik_{\alpha},L=\alpha} } \right)
        \left( \bigotimes_{\alpha \in 0, ..., L}\vec{Y}^{ik_{\alpha}} \right)
    \right] }
\end{align}
\end{widetext}
where $k_0 = j$, $n_{L + 1}' = n_L$, and $n_0' = n_1'$.

ACE can be expanded similarly using the bilinearity of the tensor product. We start from equation \ref{eqn:ace} and collapse the radial and chemical bases into one index $n$ of size $N_\text{full-basis} = S \times N_\text{basis}$ for clarity:
\begin{widetext}
\begin{align}
    B^{(\nu)}_{n_1...n_\nu} &=
    \bigotimes_{\alpha=1,...,\nu}{ A_{n_i} }
    \\ &=
    \bigotimes_{\alpha=1,...,\nu}{ \left( \sum_{k_\alpha \in \mathcal{N}(i)} R_{n_\alpha}(r_{ik_\alpha}, z_{k_\alpha})\vec{Y}^{ik_\alpha} \right) }
    \\ &= \label{eqn:ace-unwrap}
    \sum_{k_1,...,k_\nu}{\left[
        \left( \prod_{\alpha \in 1, ..., \nu} R_{n_\alpha}(r_{ik_\alpha}, z_{k_\alpha}) \right)
        \left( \bigotimes_{\alpha \in 1, ..., \nu} \vec{Y}^{ik_{\alpha}} \right)
    \right]}
\end{align}
\end{widetext}

Comparing equations \ref{eqn:unwrapped-equiv-feat} and \ref{eqn:ace-unwrap} it is immediately evident that an Allegro model with $N_\text{layer}$ layers and an ACE expansion of body order $\nu + 1 = N_\text{layer} + 2$ share the core equivariant iterated tensor products $\vec{Y}^{ij} \otimes \vec{Y}^{ik_1} \otimes ... \otimes \vec{Y}^{ik_{N_{\text{layer}}}}$. The equivariant Allegro features $\bvec{V}_{n}^{ij, L}$ are analogous --- but not equivalent --- to the full equivariant ACE basis functions $B^{(L + 1)}_{n_1...n_{L + 1}}$.\\

The comparison of these expansions of the two models emphasizes, as discussed earlier in the scaling section, that the ACE basis functions carry a full set of $n_\alpha$ indices (which label radial-chemical two-body basis functions), the number of which increases at each iteration, while the Allegro features do not exhibit this increase as a function of the number of layers. This difference is the root of the contrast between the $\mathcal{O}(N_\text{full-basis}^{\nu})$ scaling of ACE in the size of the radial-chemical basis $N_\text{full-basis}$ and the $\mathcal{O}(1)$ of Allegro. Allegro achieves this more favorable scaling through the learnable channel mixing weights.\\

A key difference between Allegro and ACE, made clear here, is their differing construction of the scalar pairwise weights. In ACE, the scalar weights carrying $ik_\alpha$ indices are the radial-chemical basis functions $R$, which are two-body functions of the distance between atoms $i$ and $k_\alpha$ and their chemistry. These correspond in Allegro to the environment embedding weights $w^{L}_{ik_\alpha,n}$, which --- critically --- are \emph{functions of all the lower-order equivariant features} $\bvec{V}_{n}^{ij, L'<L}$: the environment embedding weights at layer $L$ are a function of the scalar features from layer $L-1$ (equation \ref{eqn:env-embed-weights}) which are a function of the equivariant features from layer $L-2$ (equation \ref{eqn:latent-mlp}) and so on. As a result, the ``pairwise'' weights have a hierarchical structure and depend on all previous weights:
\begin{align}
    \label{eqn:env-weight-hierarchy}
    w_{ix, n}^L
    &= f(\bvec{V}^{ix, L-1}) \\ 
    &= f\big( \{ w^{L'}_{ix',n'} \text{  for all } n', L' < L, x' \in \mathcal{N}(i) \} \big)
\end{align}
where $f$ contains details irrelevant to conveying the existence of the dependence. We finally note that the expanded features of equation \ref{eqn:unwrapped-equiv-feat}---and thus the final features of any Allegro model---are of finite body-order if the environment embedding weights $w_{ik_\alpha n}^{L}$ are themselves of finite body-order. This condition holds if the latent and embedding MLPs are linear. If any of these MLPs contain nonlinearities whose Taylor expansions are infinite, the body-order of the environment embedding weights, and thus of the entire model, becomes infinite. Nonlinearities in the two-body MLP are not relevant to the body order and correspond to the use of a nonlinear radial basis in methods such as ACE. Allegro models whose only nonlinearities lie in the two-body embedding MLP are still highly accurate and such a model was used in the experiments on the 3BPA dataset \hyperref[sec:3bpa]{described above}.

\section{Discussion}

We introduce Allegro, a novel type of deep learning interatomic potential and demonstrate its ability to obtain highly accurate predictions of energies and forces, to scale to large atomistic systems, and to predict structural and kinetic properties from molecular dynamics simulations of complex systems. The method challenges the standard of message passing neural networks, widely used across atomistic machine learning. Our findings raise questions about the optimal choice of representation and learning algorithm for machine learning on molecules and materials. An important goal for future work is to obtain a better understanding of when explicit long-range terms are required in machine learning interatomic potentials, how to optimally incorporate them into local models, and to what extent semi-local message passing interatomic potentials may or may not implicitly capture these interactions.\\

The correspondences between the Allegro architecture and the ACE formalism discussed above also raise questions about how and why Allegro is able to outperform the systematic ACE basis expansion. We speculate that our method's performance is due in part to the learned dependence of the environment embedding weights at each layer on the full scalar latent features from all previous layers. This dependence may allow the importance of an atom to higher body-order interactions to be learned as a function of lower body-order descriptions of its environment. It stands in stark contrast to ACE, where the importance of any higher body-order interaction is learned separately from lower body-order descriptions of the local structure. We believe further efforts to understand this correspondence are a promising direction for future work.

\section{Methods}

\subsection{Software}
All experiments were run with the Allegro code available at \url{https://github.com/mir-group/allegro} under git commit \texttt{a5128c2a86350762215dad6bd8bb42875ebb06cb}. In addition, we used the NequIP code available at \url{https://github.com/mir-group/nequip} with version 0.5.3, git commit \texttt{eb6f9bca7b36162abf69ebb017049599b4ddb09c}, as well as \texttt{e3nn} with version 0.4.4 \cite{mario_geiger_2021_4735637}, PyTorch with version 1.10.0 \cite{paszke2019pytorch}, and Python with version 3.9.7. The LAMMPS experiments were run with the LAMMPS code available at \url{https://github.com/lammps/lammps.git} under git commit \texttt{9b989b186026c6fe9da354c79cc9b4e152ab03af} with the \texttt{pair\_allegro} code available at \url{https://github.com/mir-group/pair_allegro},  git commit \texttt{0161a8a8e2fe0849165de9eeae3fbb987b294079}. 
The VESTA software was used to generate Figure \ref{fig:li3po4-quench} \cite{momma2008vesta}. Matplotlib was used for plotting results \cite{Hunter:2007}.

\subsection{Reference Training Sets}
\textit{revised MD-17}: The revised MD-17 data set consists of 10 small organic molecules, for which 100,000 structures were computed at DFT (PBE/def2-SVP) accuracy using a very tight SCF convergence and very dense DFT integration grid \cite{christensen2020role}. The structures were re-computed from the original MD-17 data set \cite{chmiela2017machine, schutt2017quantum, sgdml}. The data can be obtained at \url{https://figshare.com/articles/dataset/Revised_MD17_dataset_rMD17_/12672038}. We use 950 structures for training, 50 structures for validation (both sampled randomly), and evaluate the test error on all remaining structures.\\

\textit{3BPA}: The 3BPA data set consists of 500 training structures at T=300K, and test data at 300K, 600K, and 1200K, of data set size of 1669, 2138, and 2139 structures, respectively. The data were computed using Density Functioal Theory with the $\omega$B97X exchange correlation functional and the 6-31G(d) basis set. For details, we refer the reader to \cite{kovacs2021linear}. The data set was downloaded from \url{https://pubs.acs.org/doi/full/10.1021/acs.jctc.1c00647}\\

\textit{QM9}: The QM9 data consists of 133,885 structures with up to 9 heavy elements and consisting of species H, C, N, O, F in relaxed geometries. Structures are provided together with a series of properties computed at the DFT/B3LYP/6-31G(2df,p) level of theory. The data  set was downloaded from \url{https://figshare.com/collections/Quantum_chemistry_structures_and_properties_of_134_kilo_molecules/978904}. In line with previous work, we excluded the 3,054 structures that failed the geometry consistency check, resulting in 130,831 total structures, of which we use 110,000 for training, 10,000 for validation and evaluate the test error on all remaining structures. Training was performed in units of [eV]. \\

\textit{Li\textsubscript{3}PO\textsubscript{4}}:  The Li\textsubscript{3}PO\textsubscript{4} structure consists of 192 atoms. The reference data set was obtained from two AIMD simulations both of 50 ps duration, performed in the Vienna Ab-Initio Simulation Package (VASP) \cite{vasp1,vasp2,vasp3} using a generalized gradient PBE functional \cite{PBE}, projector augmented wave pseudopotentials \cite{PAW}, a plane-wave cutoff of 400eV and a $\Gamma$-point reciprocal-space mesh. The integration was performed with a time step of 2 fs in the NVT ensemble using a Nos\'e-Hoover thermostat. The first 50 ps of the simulation were performed at T=3000K in the molten phase, followed by an instant quench to T=600K and a second 50 ps simulation at T=600K. The two trajectories were combined and the training set of 10,000 structures as well as the validation set of 1,000 were sampled randomly from the combined data set of 50,000 structures. \\

\textit{Ag}: The Ag system is created from a bulk face-centered-cubic structure with a vacancy, consisting of 71 atoms. The data were sampled from AIMD simulations at T=1,111K (90\% of the melting temperature of Ag) with Gamma-point k-sampling as computed in VASP using the PBE exchange correlation functional \cite{vasp1, vasp2, vasp3}. Frames were then extracted at least 25 fs apart, to limit correlation within the trajectory, and each frame was recalculated with converged DFT parameters. For these calculations, the Brillouin zone was sampled using a (2x2x3) Gamma-centered k-point grid, and the electron density at the Fermi-level was approximated using Methfessel-Paxton smearing \cite{methfessel1989high} with a sigma value of 0.05. A cutoff energy of 520 eV was employed, and each calculation was non-spin-polarized.

\subsection{Molecular Dynamics Simulations}

Molecular Dynamics simulations were performed in LAMMPS \cite{LAMMPS} using the pair style \texttt{pair\_allegro} implemented in the Allegro interface, available at \url{https://github.com/mir-group/pair_allegro}. We run the Li\textsubscript{3}PO\textsubscript{4} production and timing simulations under an NVT ensemble at T=600K, using a time step of 2 fs, a Nos\'e-Hoover thermostat and a temperature damping parameter of 40 time steps. The Ag timing simulations are run also in NVT, at a temperature of T=300K using a time step of 5 fs, a Nos\'e-Hoover thermostat and a temperature damping parameter of 40 time steps. The larger systems are created by replicating the original structures of 192 and 71 atoms of Li\textsubscript{3}PO\textsubscript{4} and Ag, respectively. We compute the RDF and ADFs for Li\textsubscript{3}PO\textsubscript{4}  with a maximum distance of 10 \AA \ (RDF) and 2.5 \AA \ (ADFs). We start the simulation from the first frame of the AIMD quench simulation. RDF and ADF for Allegro were averaged over 10 runs with different initial velocities, the first 10 ps of the 50 ps simulation were discarded in the RDF/ADF analysis to account for equilibration.\\

\subsection{Training details}

Models were trained on a NVIDIA V100 GPU in single-GPU training. 

\subsubsection{revMD-17 and 3BPA}

The revised MD17 models were trained with a total budget of 1,000 structures, split into 950 for training and 50 for validation. The 3BPA model was trained with a total budget of 500 structures, split into 450 for training and 50 for validation. The data set was re-shuffled after each epoch. We use 3 layers, 128 features for both even and odd irreps and a $\ell_{\text{max}}=3$. The 2-body latent MLP consists of 4 hidden layers of dimensions [128, 256, 512, 1024], using SiLU nonlinearities on the outputs of the hidden layers \cite{hendrycks2016gaussian}. The later latent MLPs consist of 3 hidden layers of dimensionality [1024, 1024, 1024] using SiLU nonlinearities for revMD-17 and no nonlinearities for 3BPA. The embedding weight projection was implemented as a single matrix multiplication without a hidden layer or a nonlinearity. The final edge energy MLP has one hidden layer of dimension 128 and again no nonlinearity. All four MLPs were initialized according to a uniform distribution of unit variance. We used a radial cutoff of 7.0 \AA\ for all molecules in the revMD-17 data set, except for Naphthalene, for which a cutoff of 9.0 \AA\ was used, and a cutoff of 5.0 \AA\ for the 3BPA data set. We use a basis of 8 non-trainable Bessel functions for the basis encoding with the polynomial envelope function using $p=6$ for revMD-17 and $p=2$ for 3BPA. RevMD-17 models were trained using a joint loss function of energies and forces: 

\begin{equation}
    \mathcal{L} = \frac{\lambda_E}{B} \sum_{b}^{B}{\left( \hat{E}_b - E_b \right)^2} +  \frac{\lambda_F}{3BN} \sum_{i=1}^{BN} \sum_{\alpha=1}^3 \lVert -\frac{\partial \hat{E}}{\partial r_{i, \alpha}}  - F_{i, \alpha} \rVert^2
\end{equation}

where $B$, $N$, $E_b$, $\hat{E}_b$, $F_{i, \alpha}$ denote the batch size, number of atoms, batch of true energies, batch of predicted energies, and the force component on atom $i$ in spatial direction $\alpha$, respectively and $\lambda_E$, $\lambda_F$ are energy and force weights. Following previous works, for the revMD-17 data the force weight was set to 1,000 and the weight on the total potential energies was set to 1. For the 3BPA molecules, as in \cite{lixin-init}, we used a per-atom MSE term that divides the energy term by $N_{atoms}^2$ because a) the potential energy is a global size-extensive property, and b) we use a MSE loss function:

\begin{equation}
    \mathcal{L} = \frac{\lambda_E}{B} \sum_{b}^{B}{\left( \frac{\hat{E}_b - E_b}{N} \right)^2} +  \frac{\lambda_F}{3BN} \sum_{i=1}^{BN} \sum_{\alpha=1}^3 \lVert -\frac{\partial \hat{E}}{\partial r_{i, \alpha}}  - F_{i, \alpha} \rVert^2
\end{equation}

After this normalization, both the energy and the force term receive a weight of 1. Models were trained with the Adam optimizer \cite{kingma2014adam} in PyTorch \cite{paszke2019pytorch}, with default parameters of $\beta_1=0.9$, $\beta_2=0.999$, and $\epsilon=10^{-8}$ without weight decay. We used a learning rate of 0.002 and a batch size of 5. The learning rate was reduced using an on-plateau scheduler based on the validation loss with a patience of 100 and a decay factor of 0.8. We use an exponential moving average with weight 0.99 to evaluate on the validation set as well as for the final model. Training was stopped when one of the following conditions was reached: a) a maximum training time of 7 days, b) a maximum number of epochs of 100,000, c) no improvement in the validation loss for 1,000 epochs, d) the learning rate dropped lower than 1e-6. We note that such long wall times are usually not required and highly accurate models can typically be obtained within a matter of hours or even minutes. All models were trained with float32 precision. 

\subsubsection{3BPA, NequIP}

The NequIP models on the 3BPA data set were trained with a total budget of 500 molecules, split into 450 for training and 50 for validation. The data set was re-shuffled after each epoch. We use 5 layers, 64 features for both even and odd irreps and a $\ell_{\text{max}}=3$. We use a radial network of 3 layers with 64 hidden neurons and SiLU nonlinearities. We further use equivariant, SiLU-based gate nonlinearities as outlined in \cite{nequip}, where even and odd scalars are not gated, but operated on directly by SiLU and tanh nonlinearities, respectively. We used a radial cutoff of 5.0 \AA\ and a non-trainable Bessel basis of size 8 for the basis encoding with a polynomial envelope function using $p=2$. We use again a per-atom MSE loss function in which both the energy and the force term receive a weight of 1. Models were trained with Adam with the AMSGrad variant in the PyTorch implementation \cite{kingma2014adam, loshchilov2017decoupled, reddi2019convergence} in PyTorch \cite{paszke2019pytorch}, with default parameters of $\beta_1=0.9$, $\beta_2=0.999$, and $\epsilon=10^{-8}$ without weight decay. We used a learning rate of 0.01 and a batch size of 5. The learning rate was reduced using an on-plateau scheduler based on the validation loss with a patience of 50 and a decay factor of 0.8. We use an exponential moving average with weight 0.99 to evaluate on the validation set as well as for the final model. Training was stopped when one of the following conditions was reached: a) a maximum training time of 7 days, b) a maximum number of epochs of 100,000, c) no improvement in the validation loss for 1,000 epochs, d) the learning rate dropped lower than 1e-6. We that such long wall times are usually not required and highly accurate models can typically be obtained within a matter of hours or even minutes. All models were trained with float32 precision. We use a per-atom shift $\mu_{Z_i}$ via the average per-atom potential energy over all training frames and a per-atom scale $\sigma_{Z_i}$ as the root-mean-square of the components of the forces over the training set.

\subsubsection{Li\textsubscript{3}PO\textsubscript{4}}

The Li\textsubscript{3}PO\textsubscript{4} model was trained with a total budget of 11,000 structures, split into 10,000 for training and 1,000 for validation. The data set was re-shuffled after each epoch. We use 1 layer, 1 feature of even parity and $\ell_{\text{max}}=1$. The 2-body latent MLP consists of 2 hidden layers of dimensions [32, 64], using SiLU nonlinearities \cite{hendrycks2016gaussian}. The later latent MLP consist of 1 hidden layer of dimensionality [64], also using a SiLU nonlinearity. The embedding weight projection was implemented as a single matrix multiplication without a hidden layer or a nonlinearity. The final edge energy MLP has one hidden layer of dimension 32 and again no nonlinearity. All four MLPs were initialized according to a uniform distribution of unit variance. We used a radial cutoff of 4.0 \AA\ and a basis of 8 non-trainable Bessel functions for the basis encoding with the polynomial envelope function using $p=48$. The model was trained using a joint loss function of energies and forces. We use again the per-atom MSE as describe above and a weighting of 1 for the force-term and 1 for the per-atom MSE term. The model was trained with the Adam optimizer \cite{kingma2014adam} in PyTorch \cite{paszke2019pytorch}, with default parameters of $\beta_1=0.9$, $\beta_2=0.999$, and $\epsilon=10^{-8}$ without weight decay. We used a learning rate of 0.001 and a batch size of 1. The learning rate was reduced using an on-plateau scheduler based on the validation loss with a patience of 25 and a decay factor of 0.5. We use an exponential moving average with weight 0.99 to evaluate on the validation set as well as for the final model. Training was stopped when one of the following conditions was reached: a) a maximum training time of 7 days, b) a maximum number of epochs of 100,000, c) no improvement in the validation loss for 1,000 epochs, d) the learning rate dropped lower than 1e-5. The model was trained with float32 precision.

\subsubsection{Ag}

The Ag model was trained with a total budget of 1,000 structures, split into 950 for training and 50 for validation. The data set was re-shuffled after each epoch. We use 1 layer, 1 feature of even parity and $\ell_{\text{max}}=1$. The 2-body latent MLP consists of 2 hidden layers of dimensions [16, 32], using SiLU nonlinearities \cite{hendrycks2016gaussian}. The later latent MLP consist of 1 hidden layer of dimensionality [32], also using a SiLU nonlinearity. The embedding weight projection was implemented as a single matrix multiplication without a hidden layer or a nonlinearity. The final edge energy MLP has one hidden layer of dimension 32 and again no nonlinearity. All four MLPs were initialized according to a uniform distribution. We used a radial cutoff of 4.0 \AA\ and a basis of 8 non-trainable Bessel functions for the basis encoding with the polynomial envelope function using $p=48$. The model was trained using a joint loss function of energies and forces. We use again the per-atom MSE as describe above and a weighting of 1 for the force-term and 1 for the per-atom MSE term. The model was trained with the Adam optimizer \cite{kingma2014adam} in PyTorch \cite{paszke2019pytorch}, with default parameters of $\beta_1=0.9$, $\beta_2=0.999$, and $\epsilon=10^{-8}$ without weight decay. We used a learning rate of 0.001 and a batch size of 1. The learning rate was reduced using an on-plateau scheduler based on the validation loss with a patience of 25 and a decay factor of 0.5. We use an exponential moving average with weight 0.99 to evaluate on the validation set as well as for the final model. The model was trained for a total of approximately 5 hours with float32 precision.

\subsubsection{QM9}

The QM9 models were trained using 110,000 structures for training and 10,000 for validation. The data set was re-shuffled after each epoch. We report two models, one with 3 layers and $\ell_{\text{max}}=2$ and another one with 1 layer and $\ell_{\text{max}}=3$, both with 256 features for both even and odd irreps. The 2-body latent MLP consists of 4 hidden layers of dimensions [128, 256, 512, 1024], using SiLU nonlinearities \cite{hendrycks2016gaussian}. The later latent MLPs consist of 3 hidden layers of dimensionality [1024, 1024, 1024], also using SiLU nonlinearities. The embedding weight projection was implemented as a single matrix multiplication without a hidden layer or a nonlinearity. The final edge energy MLP has one hidden layer of dimension 128 and again no nonlinearity. All four MLPs were initialized according to a uniform distribution. We used a radial cutoff of 10.0 \AA.\ We use a basis of 8 non-trainable Bessel functions for the basis encoding with the polynomial envelope function using $p=6$. Models were trained using a MSE loss on the energy with the Adam optimizer \cite{kingma2014adam} in PyTorch \cite{paszke2019pytorch}, with default parameters of $\beta_1=0.9$, $\beta_2=0.999$, and $\epsilon=10^{-8}$ without weight decay. In addition, we use gradient clipping by norm with a maximum norm of 100. We used a learning rate of 0.001 and a batch size of 16. The learning rate was reduced using an on-plateau scheduler based on the validation MAE of the energy with a patience of 25 and a decay factor of 0.8. We use an exponential moving average with weight 0.999 to evaluate on the validation set as well as for the final model. Training was stopped when one of the following conditions was reached: a) a maximum training time of approximately 14 days, b) a maximum number of epochs of 100,000, c) no improvement in the validation loss for 1,000 epochs, d) the learning rate dropped lower than 1e-5. All models were trained with float32 precision. Again, we note that such long wall times are not required to obtain highly accurate models. We subtract the sum of the reference atomic energies and then apply the linear fitting procedure described above using every 100-th reference label in the training set.

\section{Data Availability}

The revMD-17, 3BPA, and QM9 data sets are publicly available (see Methods section). The Li\textsubscript{3}PO\textsubscript{4} and Ag data sets will be made available upon publication. 

\section{Code Availability}

An open-source software implementation of Allegro will be available at \url{https://github.com/mir-group/allegro} together with an implementation of the LAMMPS software interface at \url{https://github.com/mir-group/pair_allegro}. 

\bibliographystyle{naturemag}
\bibliography{bib.bib}

\section{Acknowledgements}

Authors thank Dr.\ Nicola Molinari for helpful discussions.\\

Work at Harvard University was supported by Bosch Research, the US Department of Energy, Office of Basic Energy Sciences Award No. DE-SC0022199 and the Integrated Mesoscale Architectures for Sustainable Catalysis (IMASC), an Energy Frontier Research Center, Award No. DE-SC0012573 and by the NSF through the Harvard University Materials Research Science and Engineering Center Grant No. DMR-2011754. 
A.M is supported by U.S. Department of Energy, Office of Science, Office of Advanced Scientific Computing Research, Computational Science Graduate Fellowship under Award Number(s) DE-SC0021110. C.J.O. is supported by the National Science Foundation Graduate Research Fellowship Program, Grant No. DGE1745303. The authors acknowledge computing resources provided by the Harvard University FAS Division of Science Research Computing Group and by the Texas Advanced Computing Center (TACC) at The University of Texas at Austin under allocation DMR20013. This research used resources of the Argonne Leadership Computing Facility, which is a DOE Office of Science User Facility supported under Contract DE-AC02-06CH11357.\\

\section{Author Contributions}
S.B. and A.M. jointly conceived the Allegro model architecture, derived the theoretical analysis of the model, and wrote the first version of the manuscript. A.M. implemented the software and contributed to running experiments. S.B. originally proposed to work on an architecture that can capture many-body information without message passing, conducted the experiments, and contributed to the software implementation. A.J. wrote the LAMMPS interface, including parallelization across devices. L.S. proposed the linear fitting for the per-species initialization and implemented it. C.J.O. generated the Ag data. M.K. generated the AIMD data set of Li\textsubscript{3}PO\textsubscript{4}, wrote software for the analysis of MD results, and contributed to the analysis of results on this system. B.K. supervised and guided the project from conception to design of experiments, implementation, theory, as well as analysis of data. All authors contributed to the manuscript. 

\section{Competing interests}
The authors declare no competing interests.

\section{Appendix}
\subsection{A. Normalization of radial basis}
\label{apx:norm-basis}

As discussed in \hyperref[sec:normalization]{the section on normalization}, the components of our networks are designed and normalized to produce outputs of zero mean and unit variance given inputs with the same. The radial Bessel basis we use (equation \ref{eqn:radial-basis}, \cite{klicpera2020directional}) does not, as defined, satisfy this property. We found it beneficial to numerically shift and scale the basis functions to make them more closely satisfy this property. In particular we define
\begin{align}
    \mu_n &= \int_{r = 0}^{r_\text{max}}{e_{\text{RBF}, n} \mathrm{d}r} \\
    \sigma^2_n &= \int_{r = 0}^{r_\text{max}}{(e_{\text{RBF}, n} - \mu_n)^2 \mathrm{d}r } \\
    B_n(r_{ij}) &= u(r) \frac{ e_{\text{RBF}, n} - \mu_n }{\sigma_n}
\end{align}
where $e_{\text{RBF}, n}$ is the $n$th radial Bessel basis function and $u(r)$ the smooth polynomial cutoff from \cite{klicpera2020directional}. These equations follow from assuming a simple uniform distribution of $r$ between $0$ and $r_\text{max}$.

\subsection{B. Normalization of scalar residual update}
The residual update at layer $L$ is computed as a weighted sum:
\begin{equation}
    \label{eq:resnet}
    \bvec{x}^{ij, L} = \frac{1}{\sqrt{1 + \alpha^2}}\bvec{x}^{ij, L-1} + \frac{\alpha}{\sqrt{1 + \alpha^2}}\bvec{x}^{ij, L}.
\end{equation}
where $\alpha = \frac{1}{2}$ is a hyperparameter.
The forms of the coefficients in equation \ref{eq:resnet} are chosen to enforce normalization: assuming that at initialization $\bvec{x}^{ij, L-1}$ and $\bvec{x}^{ij, L}$ are negligibly correlated and each have approximately unit variance and zero mean, the residual sum will then also have approximately unit variance and zero mean.

\subsection{C. Variable per-layer cutoffs}
The Allegro formalism and implementation readily admit the use of variable cutoffs that decrease for higher layers. In particular, to use variable cutoffs, the following modifications are made:
\begin{itemize}
    \item All sums over the neighborhood of $i$, $\mathcal{N}(i)$, are restricted to be over all neighbors of $i$ within the cutoff of the current layer, $r_\text{max}^L$. In particular, the embedded environment sums only over neighbors within the current layer's cutoff.
    \item Each layer processes only edges within its cutoff. Due to the residual update, this means that the final scalar latent features of a pair are a sum over all layers for which the pair is within the layer cutoff.
    \item The smooth cutoff envelope $u(r_{ij})$ applied to the scalar latents in equation \ref{eqn:latent-mlp} is changed to a smooth cutoff envelope with the current layer's cutoff.
\end{itemize}

It is important to note that this scheme only makes sense if $r_\text{max}^{L+1} \leq r_\text{max}^L$; otherwise, the equivariant latent features required from previous layers will be missing for some edges.

\end{document}